\documentclass[ 
aps,prd,   
               preprintnumbers,numbers,sort&compress,nofootinbib,
                            showpacs,
               colorlinks,
               linkcolor=blue,   
               citecolor=blue]{revtex4-1}
   \newcommand{\exclude}[1]{}

\usepackage{graphicx,amsmath,amssymb,bm}
\usepackage{psfrag}
\usepackage{feynmp}
\usepackage{hyperref}
\usepackage{enumitem}

\newcommand{\beq}{\begin{equation}}
\newcommand{\eeq}{\end{equation}}
\newcommand{\be}{\begin{eqnarray}}
\newcommand{\ee}{\end{eqnarray}}

\def\dd{ \,\mathrm{d} }

\def\+{\dagger}
\def\la{\langle}
\def\ra{\rangle}
\def\<{\langle}
\def\>{\rangle}

\newcommand{\Lqcd}{\Lambda_{\mathrm{QCD}}}
\newcommand{\Lbar}{\Lambda_{\overline{\mathrm{QCD}}}}
\newcommand{\qcd}{{\overline{\mathrm{QCD}}}}

\begin{document}


\title{Inflaton as an auxiliary topological field in a QCD-like system.}

\author{   Ariel R. Zhitnitsky} 
 \affiliation{Department of Physics \& Astronomy, University of British Columbia, Vancouver, B.C. V6T 1Z1, Canada}


\begin{abstract}
We propose a new scenario for early cosmology, where inflationary de Sitter phase  dynamically occurs. 
 The effect emerges as a result of   dynamics of the  topologically  nontrivial  sectors in expanding universe. 
 Technically the effect  can be described in terms of the  auxiliary   fields which effectively describe the dynamics  of the topological sectors in a gauge theory. Inflaton in this framework is an auxiliary topological non-propagating field with no canonical kinetic term, similar to known topologically ordered phases in condensed matter systems. We explain many deep questions in this framework using the so-called  weakly coupled ``deformed QCD" toy model.
 While this  theory is weakly coupled gauge theory, it  preserves all the crucial elements of strongly interacting gauge theory, including confinement, nontrivial $\theta$ dependence, degeneracy of the topological sectors, etc.   We discuss  a    specific   realization of these ideas using   a scaled up version of QCD, coined as $\qcd$, with the scale $M_{PL}\gg \Lbar\gg    \sqrt[3]{M_{EW}^2M_{PL}}\sim 10^8~ {\mathrm{GeV}}$.  If no other fields are present in the system
 de Sitter phase will be  the final destination of evolution of the universe. If other interactions are present in the system, the inflationary de Sitter phase lasts  for a finite period of time. The inflation starts  from the thermal equilibrium state long  after the $\qcd$ -confinement phase transition at temperature $T_{i}\sim \Lbar\sqrt{\frac{\Lbar}{M_{PL}}}$. The end of inflation   is triggered by the  coupling with   gauge bosons from the Standard Model. 
The corresponding interaction is unambiguously fixed by the triangle anomaly.  Number of e-folds in the $\qcd$-inflation framework   is determined by the gauge  coupling constant at the moment of inflation, and estimated as $N_{\rm inf}\sim \alpha_s^{-2}\sim 10^2$.
 
 \end{abstract}

\maketitle

\section{Introduction. Motivation.}
It is well known that the deep issue inflation addresses (among many other things) is the origin of the large-scale homogeneity of the observable universe\cite{inflation, linde,mukhanov}. The crucial element of this idea is to have a period of  evolution of the universe which can be 
   well approximated by the de Sitter   behaviour. In this case the scale parameter ${\rm{a}} (t)$ and 
the equation of state takes the following  approximate  form, 
 \be
\label{a}
{\rm{a}}(t)\sim \exp (Ht), ~~~ \epsilon\approx  -p.
\ee
It is normally assumed that such equation of state can be achieved in quantum field theory (QFT) by  
  assuming the existence of a scalar matter field  $\Phi$ with a non-vanishing potential energy density $V(\Phi)$. The shape of this potential energy can be adjusted in a such a way that the  contribution to energy density and pressure   is in agreement with the above equation of state. The inflationary scenario can be described in a simplified way   as follows: at the initial time, the scalar field $\Phi$ is displaced from the minimum of its potential. Since the potential $V(\Phi)$ is tuned to be very flat, the scalar field motion is very slow. Therefore, the scalar field potential energy density remains almost constant, whereas all other forms of matter redshift. Thus, at some time $t_i$, the scalar field potential energy starts to dominate and inflation begins. Once the scalar field has decreased to a critical value which in many models is close to the Planck scale $M_{PL}$, the scalar field kinetic energy begins to dominate over the potential energy and inflation ends at time $t_r$. There are many problems with this picture. We shall not   address those problems in the present work  by referring to  review articles \cite{Brandenberger:2011gk,Brandenberger:2012uj}, see also very recent papers \cite{Ijjas:2013vea,Ijjas:2014nta} which address   the problems the inflation paradigm faces after Planck2013. The only element which is crucial for the entire framework outlined above, is   merely existence of a new dynamical degrees of freedom, the {\it inflaton} which is typically approximated by a scalar field $\Phi$, while its dynamics is governed by the potential $V(\Phi)$. Even in string inspired models the presence of such dynamical field 
  seems unavoidable. For example, in the so-called KKLT construction \cite{KKLT} the  inflaton field is associated with one of the moduli, see \cite{linde,mukhanov} with overview of many other models.
  
  In the present work we advocate a fundamentally different view on the nature and origin of the inflaton field.
  We shall argue that the role of the inflaton may play an auxiliary topological field which normally emerges in description of a topologically ordered gauge system.  
  These fields do not propagate, they do not have canonical kinetic terms, as they are auxiliary fields effectively describing the dynamics of the topological sectors which are present in the system. Nevertheless, the effects which are described in terms of these topological fields are quite physical and very real as we shall argue in the present work.
  
  The gauge theory which has all the features required to describe the inflationary phase in evolution of the universe is very much the same as strongly coupled QCD  but with drastically different scale, much larger than conventional $ \Lqcd$.
  Many  relevant elements which are required for  the inflationry phase to be operational are in fact have been tested using the numerical lattice Monte Carlo simulations. 
   However, in order to  study  some 
  deep physical properties  of the system  related to the large distance behaviour we formulate a simplified version    of QCD, the so-called ``deformed QCD" which is a weakly coupled gauge theory, but nevertheless preserves all the crucial elements of strongly interacting QCD, including confinement, nontrivial $\theta$ dependence, degeneracy of the topological sectors, etc. The emergence of the dynamical inflationary phase is much easy to explain using the analytically tractable  ``deformed QCD" model, 
  rather than referring to some specific numerical results. 
  
  For impatient readers   we formulate here the basic findings of our studies. The key element for  our work is the presence of the degenerate topological sectors   in the gauge theory denoted as $\qcd$ with the scale $\Lbar\gg \Lqcd$. The dynamics of these pure gauge sectors in the gravitational expanding background  can be formulated in terms of an auxiliary topological field. The relevant dynamics of this auxiliary field  precisely represents  the physics which is normally attributed to the inflaton.  $\qcd$ is an asymptotically free gauge theory such that the UV completion requirement is obviously satisfied in the model.   The inflation with almost de Sitter behaviour ${\rm{a}}(t)\sim \exp (Ht)$ starts  from the thermal equilibrium state at temperature $T_{i}\sim \Lbar\sqrt{\frac{\Lbar}{M_{PL}}}$ long  after the $\qcd$ -confinement phase transition at $T_c\sim \Lbar$. The   inflation ends as a result of interaction  of the  $\qcd$ fields with   gauge bosons from the Standard Model.

  An educated  reader may immediately get suspicious with a question:  how does a gapped theory with typical fluctuations $r\sim 1/\Lbar$ could ever influence the physics with vastly different scale $r\sim 1/H$ where $H$ is the Hubble expansion rate at the time $T_i$.   One of the main objectives of the present work is precisely to address  this question using  a weakly coupled ``deformed QCD" where computations can be performed in theoretically controllable way. A short answer on this question is that $\qcd$ behaves similar to a topologically ordered  condensed matter (CM) system which  is normally  gapped but still  remains  highly  sensitive to arbitrary large distances.
  
   It might be instructive to get  some intuitive picture for the {\it inflaton} in this framework formulated in terms of a CM analogy before we proceed with formal computations. Imagine,  we study 
    the  Aharonov -Casher effect.  We insert an external charge into superconductor when the electric field is exponentially suppressed $\sim \exp(-r/\lambda)$ with $\lambda $ being the penetration depth. Nevertheless,  a neutral magnetic fluxon will be still sensitive to an inserted external charge at arbitrary large distance in spite of the screening of the physical field. The effect is pure topological and non-local in nature
    and can be explained in terms of pure gauge sectors which are responsible for this long range dynamics. Imagine now that we study the same effect but in expanding  universe. The corresponding topological sectors will be modified due to the external background. However, this modification can not be described in terms of any dynamical fields, as there are no any propagating long range fields in the system as physical electric field is screened. For this simplified example the dynamics of the {\it inflaton } corresponds    to the effective description of the   topological sectors  variation when the background changes. 
    The effect is obviously non-local in nature as   the  Aharonov -Casher effect itself  is a non-local phenomenon. Furthermore, the effect can not be formulated in terms of any physical propagating degrees of freedom (such as $\Phi$ field mentioned above) as pure gauge, but topologically nontrivial,  configurations can  not  be described in terms of a local  physical propagating field $\Phi$. We elaborate on this analogy in a much more precise and specific way in Appendix \ref{H} of this work. 
    
    Our presentation is organized as follows.
    In section \ref{deformedqcd} we   overview the weakly coupled ``deformed QCD" model. In section \ref{BF-section} we use this model to describe the relevant  effects in terms of auxiliary non-propagating topological   fields. In particular, we discuss 
    the non-dispersive  $\theta$ dependent   contribution to the energy $E_{\mathrm{vac}}(\theta)$  which can not be expressed in terms of  any physical propagating degrees of freedom, see below.  As this  contact term   can not be associated with any physical fields we coin this type of energy as  the {\it ``strange energy"} in this paper.  We also discuss how this    {\it ``strange energy"}  varies 
    when the background changes.  The basic idea of sections  \ref{deformedqcd} and   \ref{BF-section} 
    is to reveal  some very deep properties of the {\it ``strange energy"}  using a simplified model.      Many of these properties are well studied  using the numerical lattice computations in strongly coupled QCD, but it is very instructive to understand  these fundamental features  using some analytical methods in a simplified model.
         In section \ref{inflation} we assume that the physics in strongly coupled 
    $\qcd$ is very much the same as in the weakly coupled ``deformed QCD" model. With this assumption we  demonstrate the emergence of the de Sitter like behaviour in expanding universe when scale parameter ${\rm a}(t)$ shows  the exponential growth, ${\rm a}(t)\sim \exp (Ht)$. In section \ref{reheating} we sketch our vision  of the reheating epoch, and explain how it could in principle emerge in this framework.
       We conclude in section \ref{conclusion} with a large number of questions and problems for future studies within this new framework, which we call the {\it $\qcd$ - inflation}. In particular, we comment on how this 
       fundamentally new type of ``strange energy" can be, in principle,  studied  in a   terrestrial table-type  laboratory experiment by measuring some specific  corrections to the observed Casimir forces. 
    
  \section{The nature of the ``Strange" energy    in the deformed QCD model} \label{deformedqcd}

The goal here is to  overview  the    deformed QCD model  where the relevant dynamics 
describing the ``strange"  vacuum energy can be explicitly seen and studied. This theory is weakly coupled gauge theory, 
but nevertheless preserves all the crucial elements of strongly interacting QCD, including confinement, nontrivial $\theta$  dependence, degeneracy of the topological sectors, etc. Furthermore, it has been claimed  \cite {Shifman:2008ja,Yaffe:2008} that there is no any phase transition in passage from weakly coupled deformed QCD to strongly coupled QCD. Crucial element for this work is presence of the contact non-dispersive term in topological susceptibility, see below,  which can not be associated with any physical propagating degrees of freedom. 
As this contribution  to the  $\theta$ dependent portion of the energy $E_{\mathrm{vac}}(\theta)$ is the key element in our discussions 
in the present work, we specifically concentrate on the nature and origin of this term. Precisely this   energy  which can not be expressed in terms of real propagating degrees of freedom will be the source of the {\it $\qcd$ -inflation} as we argue in section \ref{inflation}.


We start with pure Yang-Mills (gluodynamics) with gauge group $SU(N)$ on the manifold $\mathbb{R}^{3} \times S^{1}$ with the standard action
\be \label{standardYM}
	S^{YM} = \int_{\mathbb{R}^{3} \times S^{1}} d^{4}x\; \frac{1}{2 g^2} \mathrm{tr} \left[ F_{\mu\nu}^{2} (x) \right],
\ee
and add to it a deformation action  \cite {Shifman:2008ja,Yaffe:2008},
\be \label{deformation}
	\Delta S \equiv \int_{\mathbb{R}^{3}}d^{3}x \; \frac{1}{L^{3}} P \left[ \Omega(\mathbf{x}) \right],
\ee 
built out of the Wilson loop (Polyakov loop) wrapping the compact dimension
\be \label{loop}
	\Omega(\mathbf{x}) \equiv \mathcal{P} \left[ e^{i \oint dx_{4} \; A_{4} (\mathbf{x},x_{4})} \right].
\ee
The parameter  $L$ here  is the length of the compactified dimension
which is assumed to be small. 
 The coefficients of the polynomial  $ P \left[ \Omega(\mathbf{x}) \right]$ can be suitably chosen such that the deformation potential (\ref{deformation}) forces unbroken symmetry at any compactification scales. At small compactification $L$ the gauge coupling  is small so that 
the semiclassical computations are under complete theoretical control \cite {Shifman:2008ja,Yaffe:2008}.

As described in \cite {Shifman:2008ja,Yaffe:2008}, the proper infrared description of the theory is a dilute gas of $N$ types of monopoles, characterized by their magnetic charges, which are proportional to the simple roots and affine root $\alpha_{a} \in \Delta_{\mathrm{aff}}$ of the Lie algebra for the gauge group $U(1)^{N}$. 
 For a fundamental monopole with magnetic charge $\alpha_{a} \in \Delta_{\mathrm{aff}}$, the topological charge $Q$ and the Yang-Mills action $S_{YM}$ are given by 
\be \label{topologicalcharge}
	Q = \int_{\mathbb{R}^{3} \times S^{1}} d^{4}x \; \frac{1}{16 \pi^{2}} \mathrm{tr} \left[ F_{\mu\nu} \tilde{F}^{\mu\nu} \right]
		= \pm\frac{1}{N},~~~~~~~
	S_{YM} = \int_{\mathbb{R}^{3} \times S^{1}} d^{4}x \; \frac{1}{2 g^{2}} \mathrm{tr} \left[ F_{\mu\nu}^{2} \right]= \frac{8 \pi^{2}}{g^{2}} \left| Q \right|.
		 \ee
 The  $\theta$-parameter in the Yang-Mills action can be included in conventional way,
\be \label{thetaincluded}
	S_{\mathrm{YM}} \rightarrow S_{\mathrm{YM}} + i \theta \int_{\mathbb{R}^{3} \times S^{1}} d^{4}x\frac{1}{16 \pi^{2}} \mathrm{tr}
		\left[ F_{\mu\nu} \tilde{F}^{\mu\nu} \right],
\ee
with $\tilde{F}^{\mu\nu} \equiv \epsilon^{\mu\nu\rho\sigma} F_{\rho\sigma}$.

The partition function for the system of interacting monopoles, including $\theta$ parameter, can be represented in the dual sine-Gordon form as follows \cite{Shifman:2008ja,Yaffe:2008,Thomas:2011ee},
\be
\label{thetaaction}
{\cal Z} [\bm{\sigma}]\sim\int {\cal{D}}[\bm{\sigma}]e^{-S_{\mathrm{dual}}[\bm{\sigma}]}, ~~~
	S_{\mathrm{dual}} [\bm{\sigma}]= \int_{\mathbb{R}^{3}}  d^{3}x \frac{1}{2 L} \left( \frac{g}{2 \pi} \right)^{2}
		\left( \nabla \bm{\sigma} \right)^{2}  - \zeta  \int_{\mathbb{R}^{3}}  d^{3}x \sum_{a = 1}^{N} \cos \left( \alpha_{a} \cdot \bm{\sigma}
		+ \frac{\theta}{N} \right)  	, 
\ee
where $\zeta$ is magnetic monopole fugacity which can be explicitly computed in this model using the conventional semiclassical approximation. The $\theta$ parameter enters the effective Lagrangian (\ref{thetaaction}) as $\theta/N$ which is the direct consequence of the fractional topological charges of the monopoles (\ref{topologicalcharge}). Nevertheless, the 
theory is still $2\pi$ periodic. This
  $2\pi$ periodicity of the theory is restored not due to the $2\pi$ periodicity of Lagrangian (\ref{thetaaction}).
  Rather, it is restored as a result of   summation over all branches of the theory when the  levels cross at
   $\theta=\pi (mod ~2\pi)$ and one branch replaces another and becomes the lowest energy state as discussed in \cite{Thomas:2011ee}.
 Finally, the vacuum energy density of the system $E_{\mathrm{vac}}(\theta)$  follows from (\ref{thetaaction}) and is given by
   \be \label{E_vac}
	E_{\mathrm{vac}}(\theta) =-\frac{N\zeta}{L}\cos\left(\frac{\theta}{N} \right),  
	  \ee
  where $|\theta|<\pi$ corresponds to the first branch. We should note that  the $\theta$ parameter is assumed to be zero in this work. 
  Nevertheless, we keep  $\theta$ parameter explicitly  in some formulae below  because it allows us to reconstruct many important and exact relations such as the couplings to other fields.  To avoid any confusions which may occur  from appearance the $\theta $ parameter in some  formulae in this work we should emphasize that   the  $\theta$  it is not a dynamical variable in this work such that the axion field  is not present in this system.
    
    Our goal now is to understand the nature of this $\theta$- dependent portion of the vacuum energy (\ref{E_vac}) as it   plays a key role in our discussions in next sections.
  As we shall argue  below  this energy is very different from conventional energy normally attributed to physical states. In fact,  the vacuum energy
  $E_{\mathrm{vac}}(\theta)$ as we shall discuss below 
  can not be associated with any physical propagating degrees of freedom.  Before we demonstrate this unusual feature, we have to make a short detour. 
    
  We start our  short detour with overview on  formulation and resolution of the so-called $U(1)_A$ problem in strongly coupled QCD~\cite{witten,ven,vendiv} which is ultimately related to the ``strange" nature of the vacuum energy (\ref{E_vac}). We 
   introduce  the topological susceptibility $\chi$
 which plays a crucial role in resolution of the   $U(1)_A$ problem as follows\footnote{We use the Euclidean notations  where  path integral computations are normally performed.}
\be
\label{chi}
 \chi (\theta =0) =    \left. \frac{\partial^2E_{\mathrm{vac}}(\theta)}{\partial \theta^2} \right|_{\theta=0}= \lim_{k\rightarrow 0} \int \!\dd^4x e^{ikx} \la T\{q(x), q(0)\}\ra  ,
 \ee
where     $\theta$ parameter   enters the  Lagrangian (\ref{thetaincluded}) along with  topological density operator $q (x)=
\frac{1}{16 \pi^{2}} \mathrm{tr}[ F_{\mu\nu} \tilde{F}^{\mu\nu}]$ and $E_{\mathrm{vac}}(\theta)$ is the vacuum energy density
which can be explicitly computed in the deformed QCD model (\ref{E_vac}).       
It is important  that the topological susceptibility $\chi$  does not vanish in spite of the fact that $q= \partial_{\mu}K^{\mu}$ is total divergence. This feature is very different from any conventional correlation functions  which normally must  vanish  at zero momentum if the  corresponding operator  can be represented as 
total divergence.  Furthermore, any physical $|n\ra$   state gives a negative contribution to this 
diagonal correlation function
\be	\label{G}
  \chi_{\rm dispersive} \sim  \lim_{k\rightarrow 0} \int d^4x e^{ikx} \la T\{q(x), q(0)\}\ra  
  \sim 
    \lim_{k\rightarrow 0}  \sum_n \frac{\la  0 |q|n\ra \la n| q| 0\ra }{-k^2-m_n^2}\simeq -\sum_n\frac{|c_n|^2}{m_n^2} \leq 0,  
\ee
 where   $m_n$ is the mass of a physical $|n\ra$ state,  $k\rightarrow 0$  is  its momentum, and $\la 0| q| n\ra= c_n$ is its coupling to topological density operator $q (x)$.
 At the same time the resolution of the $U(1)_A$ problem requires a positive sign for the topological susceptibility (\ref{chi}), see the original reference~\cite{vendiv} for a thorough discussion, 
\be	\label{top1}
  \chi_{\rm non-dispersive}= \lim_{k\rightarrow 0} \int \!\dd^4x e^{ikx} \la T\{q(x), q(0)\}\ra > 0.~~~
\ee
Therefore, there must be a contact contribution to $\chi$, which is not related to any propagating  physical degrees of freedom,  and it must have the ``wrong" sign. The ``wrong" sign in this paper implies a sign 
  which is opposite to any contributions related to the  physical propagating degrees of freedom (\ref{G}). 
  The ``strange energy" in this paper implies the $\theta$ dependent portion of the energy (\ref{chi}) which {\it can not} be formulated  in terms of conventional propagating degrees of freedom as it has pure non-dispersive nature according to eqs. (\ref{G}, \ref{top1}).

    In the framework \cite{witten} the contact term with ``wrong" sign  has been simply postulated, while in refs.\cite{ven,vendiv} the Veneziano ghost (with a ``wrong" kinetic term) had been introduced into the theory to saturate the required property (\ref{top1}).   
  Furthermore, as we discuss below the  contact term has  the structure $\chi \sim \int d^4x \delta^4 (x)$.
  The significance of this structure is  that the gauge variant correlation function in momentum space
  \be
  \label{K}
   \lim_{k\rightarrow 0} \int d^4x e^{ikx} \la K_{\mu}(x) , K_{\nu}(0)\ra\sim   \frac{k_{\mu}k_{\nu}}{k^4}
   \ee 
  develops  a topologically protected  ``unphysical" pole which does not correspond to any propagating massless degrees of freedom, but nevertheless must be present in the system. Furthermore, the residue of this   pole has the ``wrong sign", which  precisely corresponds to the Veneziano ghost  contribution saturating the non-dispersive term  in gauge invariant correlation function (\ref{top1}),
   \be
  \label{K1}
   \< q({x}) q({0}) \> \sim  \la \partial_{\mu}K^{\mu}(x) , \partial_{\nu}K^{\nu}(0)\ra \sim \delta^4(x)
   \ee 
   We conclude this short detour with the following remark. The entire framework, including the singular behaviour of
  $ \< q({x}) q({0}) \>$   with the ``wrong sign",  has been well confirmed by numerous  lattice simulations in strong coupling regime, and it is accepted by the community as a standard resolution of the $U(1)_A$ problem.  Furthermore, it has been argued long ago in ref.\cite{Luscher:1978rn}
  that the gauge theories may exhibit the ``secret long range forces" expressed in terms of the correlation function (\ref{K}).

     We now return to the deformed QCD model where every single question (including the  non-dispersive  nature of ``strange energy") can be answered as we are dealing with the weakly coupled gauge theory.  The study of this object precisely shows how the non-dispersive  vacuum energy  (i.e. not related to any propagating degrees of freedom)      may emerge in the system. As we shall argue  in section \ref{inflation} precisely this type of energy, which is fundamentally not describable  in terms of physical propagating degrees of freedom,  may be  responsible for the {\it $\qcd$ inflation} .
        
 The topological susceptibility in the deformed QCD   model 
 can be explicitly computed as it is saturated by fractionally charged weakly interacting monopoles, and it is given by \cite{Thomas:2011ee}
\be \label{YM}
  \chi_{YM} = \int d^4 x \< q(\bold{x}) , q(\bold{0}) \>	=\frac{\zeta}{NL} \int d^3 x \left[ \delta(\bold{x}) \right], 
\ee
which precisely corresponds to the vacuum energy (\ref{E_vac}) in this model after differentiation with respect to $\theta$ parameter according to (\ref{chi}).
The topological susceptibility  has the required  ``wrong sign" as this  contribution is not related to any physical propagating degrees of freedom, and it has a 
$\delta(\bold{x})$ function structure which implies the presence of the pole (\ref{K}). However, there are no any physical massless states in the system as  it  is gapped, and  the computations \cite{Thomas:2011ee} leading to (\ref{YM}) are accomplished without any topological or any other unphysical degrees of freedom. 
Instead, this term is described in terms of the tunnelling events between  different (but physically equivalent) topological sectors in the system. 
The monopoles in this framework are not real particles, they are pseudo-particles which live in Euclidean space and describe the physical tunnelling processes between different topological sectors $|k\ra$ and $| k+1 \ra$. The ``strange" energy of the system (\ref{E_vac})   should be interpreted  as number of tunnelling events per unit time $L$  per unit volume $V$
\be
\label{zeta}
  \left(\frac{  {\rm number ~of  ~tunnelling ~ events}}{VL}\right)=\frac{N\zeta}{L},~~~~~~~~~~ E_{\mathrm{vac}}
 =- \frac{N\zeta}{L},
\ee
where $\zeta$ is the monopole fugacity to be understood as a number of tunnelling events for a given type of monopole per unit time $L$. There are $N$ different types of monopoles which explains the normalization in eq.(\ref{zeta}). 
Precisely this interpretation reveals   the non-dispersive nature of this ``strange" energy which can not be attributed to any physical propagating degrees of freedom. It is quite obvious that the nature of this ``strange" energy is very different from conventional vacuum energy formulated in terms of a dynamical scalar field $\Phi$, such as the Higgs field which is the key player of  the standard model, or conventional inflaton field which is the key player of the inflation formulated  in terms of a dynamical $\Phi$ field~\cite{linde,mukhanov}. 

From the discussions presented above it must be obvious that this ``strange" energy which eventually will be responsible for the {\it $\qcd$- inflation} has non-dispersive nature, i.e. can not be associated with any physical propagating degrees of freedom. 
Furthermore, the ``strange" energy  can not be seen at any level in perturbation theory as $\zeta\sim \exp(-1/g^2)$. 
Finally, the generation of this  ``strange" energy can be thought as a non-local  phenomenon as the tunnelling events which are responsible for $E_{\mathrm{vac}}$ are formulated in terms of the    
transitions between distinct  topological sectors $| k\ra$.  At the same time, these $| k\ra$ sectors are constructed by using  the  large gauge transformation operator $\cal{T}$ which itself is a non-local operator, see Appendix \ref{meaning} for the details.  Nevertheless, as we shall argue below,  
this ``strange" vacuum energy (\ref{zeta}) is finite and uniquely defined and can not be removed from the system by any  subtractions    or redefinitions of  the observables. The arguments are based on exact Ward Identities, see next section \ref{BF-section}. One should also note that all these unusual features have been well studied  in strongly coupled QCD  using the  lattice numerical simulations, see e.g.  \cite{Zhitnitsky:2013hs} with large number of  references on the original lattice results.

 In the next section  we shall  re-derive  the same $\delta (\mathbf{x})$ function  (\ref{YM}) 
 in terms of an auxiliary   topological   field  for deformed QCD. This will further illuminate the IR nature of the contact term. It will also reveal  the nature   of the topologically protected massless pole (\ref{K}) which results from the dynamics of  an auxiliary  topological field. Precisely  this field will play a key role in our discussions  of the  {\it $\qcd$ inflation} in section \ref{inflation} as its dynamics   will be determined by the evolution of   the non-propagating auxiliary topological field.

\section{Topological auxiliary field as an  inflaton} \label{BF-section}
The goal of this section is to express the same ``strange" vacuum  energy (\ref{zeta}) in term of a quantum  field which  accounts  for the physics of tunnelling transitions discussed above. 
We should emphasize that the reformulation of the same physics in terms of a quantum  field rather than in terms of explicit 
computation of the partition function by summing over all topological sectors  is not a mandatory procedure, but a matter of convenience. Similarly, the description of a topologically ordered phase in condensed matter physics in terms of Chern Simons effective Lagrangian is a matter of convenience rather than a necessity. 
The same comment also applies to our case when  an auxiliary, not dynamical, topological field   (effectively describing the dynamics of the topological sectors), the    {\it inflaton},  is introduced for the  great convenience as we shall see in a moment.

When the same physics is reformulated in terms of  a quantum  field all the unusual features discussed  above will be much easy to understand. 
Furthermore,  
the corresponding reformulation  of the system in terms of a quantum field will be extremely useful 
in  addressing the question on possible changes of the ``strange" enenrgy when the background varies, see  section \ref{odd} and Appendix \ref{H}. In addition, the  reformulation  of the system in terms of a quantum field is the key element 
in formulation  of the problem of the {\it reheating} epoch within the  {\it $\qcd$ inflation} framework, see   section \ref{reheating}.  Finally, one should keep in mind that the  {\it inflaton} is an emergent field: it only appears  in the confined $\qcd$ phase, while in the  deconfined phase it does not appear in the system. This simple comment will have, in fact, some profound observational consequences when   one  compares  the  {\it $\qcd$- inflaton}   with conventional   $\Phi$- inflaton  which always existed in the system, see section \ref{conclusion}   for the related  discussions.

\subsection{ Topological auxiliary field as a source of the ``strange" energy}
The basic idea to describe the IR physics in terms of an auxiliary field is to insert the corresponding $\delta$- function into the path integral with a Lagrange multiplier and integrate out the fast degrees of freedom while keeping the slow degrees of freedom which are precisely the  auxiliary fields. This formal trick is widely used in particle physics and condensed matter (CM) physics.  In particular, it is extremely useful in description of the topologically ordered phases when the IR physics is formulated in terms of the  topological Chern-Simons (CS) like Lagrangian, see e.g. \cite{BF} and references therein. One  should emphasize that the corresponding 
CM physics, such as calculation  of the braiding phases between quasiparticles, computation of the degeneracy etc,  can be computed (and in fact originally had been computed)  without Chern-Simons Lagrangian and without auxiliary fields. Nevertheless, the discussions of the IR physics in terms of CS like effective action  is proven to be very useful, beautiful and beneficial. 

 For the deformed QCD model the corresponding computations have been carried out in \cite{Zhitnitsky:2013hs} where it has been demonstrated that all unusual properties of the ``strange" energy (\ref{zeta}) including its non-dispersive nature can be formulated in terms of auxiliary long range topological $a(\mathbf{x}), b(\mathbf{x})$ fields with the action
 \be
\label{b-action}
{\cal Z} [\bm{\sigma},   b, a]&\sim&\int {\cal{D}}[b]{\cal{D}}[\bm{\sigma}]{\cal{D}}[ a]e^{-S_{\rm top}[b, a]-S_{\rm dual}[\bm{\sigma},   b]}, ~~~~~~
S_{\rm top}[b, a] =
 \frac{-i }{4 \pi N}  \int_{\mathbb{R}^{3}}  d^{3}x     b(\mathbf{x})\vec{\nabla}^2 a (\mathbf{x}) ,
  \\
	S_{\rm dual}[\bm{\sigma},   b] &=& \int_{\mathbb{R}^{3}}  d^{3}x \frac{1}{2 L} \left( \frac{g}{2 \pi} \right)^{2}
		\left( \nabla \bm{\sigma} \right)^{2} \nonumber - \zeta  \int_{\mathbb{R}^{3}}  d^{3}x \sum_{a = 1}^{N} \cos \left( \alpha_{a} \cdot \bm{\sigma}
		+ \frac{\theta-b(\mathbf{x})}{N} \right).
		 \nonumber
\ee
   In this formula the topological action $S_{\rm top}[b, a] $ can be expressed  as a conventional CS effective Lagrangian  \cite{Zhitnitsky:2013hs} but in this  work we keep  only relevant for the present discussions components
   represented by the scalar $b(\mathbf{x}), a(\mathbf{x}) $ fields.  
   
   Now we can compute the ``strange" energy 
   which has the physical meaning of number of the tunnelling events per unit volume per unit time (\ref{zeta}) 
   in terms of the auxiliary fields. The corresponding formula can be represented  in terms of  the   correlation function $ \la \vec{\nabla}^2 a (\mathbf{x}),  \vec{\nabla}^2 a (\mathbf{0})\ra$  as follows 
    \be
\label{YM_top}
E_{\rm vac}= -N^2\lim_{k\rightarrow 0} \int d^4x e^{ikx} \la q(\mathbf{x}), q(\mathbf{0})\ra =-\frac{N\zeta}{L}\int d^3x \delta^3(\mathbf{x})= -\frac{N\zeta}{L},~~ {\rm where}~~~ q(\mathbf{x})=\frac{-1 }{4 \pi NL}    \vec{\nabla}^2 a (\mathbf{x}). 
\ee
We obviously reproduce our previous result (\ref{YM}, \ref{zeta}), but now it is formulated in terms of the long -ranged auxiliary topological fields. We emphasize again: we have not introduced any new degrees of freedom into the system. The fluctuating  $b(\mathbf{x}), a(\mathbf{x}) $ fields simply reflect the long distance dynamics of the degenerate topological sectors which exist independently from our description in terms of  $b(\mathbf{x}), a(\mathbf{x}) $ fields. However, in previous computations  (\ref{YM}, \ref{zeta}) we had to sum over all monopoles, their positions, interactions  and orientations. Now this problem is simplified as it is reduced to the computation of the correlation function constructed from the auxiliary fields governed by the action (\ref{b-action}).  

We shall argue in section \ref{inflation}  that  the ``strange"   energy  (\ref{YM}, \ref{zeta}) can serve as the vacuum energy during the   inflation period  in expanding universe. Therefore,  we identify the corresponding auxiliary 
$[a(\mathbf{x}), b(\mathbf{x})] $ fields  which saturate this energy (\ref{YM_top}) with {\it inflaton} in this model.
  We emphasize again that the corresponding dynamics can not be formulated in terms of a
 canonical scalar field $\Phi$ with any local potential $V(\Phi)$ as it is known that the   dynamics governed by CS-like action is truly non-local.   There is a large number of CM systems (realized in nature)  where CS action plays a key role with explicit manifestation of the  non-locality in the system.   It has been also argued that the deformed QCD model which is explored in this section also belongs to a topologically ordered phase   with many  features which normally accompany the topological phases~\cite{Zhitnitsky:2013hs}. 
 What is important  is that  the auxiliary 
$[a(\mathbf{x}), b(\mathbf{x})] $ fields  emerge in the system only in confined phase. In the deconfined phase the
``strange energy"  (\ref{YM},\ref{zeta}) vanishes because the topological susceptibility vanishes in deconfined phase. This is in huge contrast with conventional inflaton $\Phi$ field which always existed in the system. 
 
 What is the physical meaning of this auxiliary field $a(\mathbf{x}) $ which we identify with {\it inflaton}?
 What is the best way  to  visualize it  on the intuitive level? 
 From our construction one can easily see that while $a(\mathbf{x})$ does not carry a colour index. Still,  it  is not a colour singlet as it has nontrivial  transformation 
 properties under large gauge transformation  \cite{Zhitnitsky:2013hs}. In fact   
 our field $ \nabla_i a(\mathbf{x}) $ transforms  as the $K_{i} (\mathbf{x})$. One should not confuse $a(\mathbf{x})$ field with magnetic potential in this model. The physical magnetic potential is characterized in this model by roots $\alpha_{a} \in \Delta_{\mathrm{aff}}$ of the Lie algebra in contrast with transformation properties of $a(\mathbf{x})$ field which essentially represents a longitudinal portion of $K_{i} (\mathbf{x})$.  
 The best intuitive way to think about $a(\mathbf{x})$ field is to imagine a coherent superposition (of non-abelian gauge fields) which has  nontrivial   properties under   large gauge transformations. This superposition   is precisely represented by the longitudinal component of $K_i$ operator.  What is the physical meaning of   $b(\mathbf{x}) $ field? As we discuss in section \ref{reheating} this field can be thought as an external axion $\theta(x)$ field, without kinetic term, though.  
 
 The vacuum energy of the system computed in terms  of the   $a(\mathbf{x})$ field  is given by eq.(\ref{YM_top}).
 Is this energy physically observable? Our ultimate answer is ``yes" as we cannot 
   redefine the energy-momentum operator   to remove this ``strange" energy (\ref{zeta},\ref{YM_top}) from the system\footnote{In fact, one can argue that the  generation of the ``strange" energy    is not the only manifestation of the topological sectors in the gauge theory. A similar  contribution with a ``wrong sign" which is not related to any physical degrees of freedom was previously observed in computations of black hole entropy \cite{Kabat:1995eq}. The corresponding contact term from \cite{Kabat:1995eq} 
   leads to the well known mismatch between computations of the black hole entropy and entropy of entanglement for vector gauge fields.  It has been conjectured in \cite{Zhitnitsky:2011tr} that this mismatch is the consequence of the same topological sectors  of the gauge theories 
   which is the subject of the present work. In fact, this conjecture is supported in some way by computations in a simplified model \cite{Donnelly:2012st}, see also \cite{Solodukhin:2012jh,Zhitnitsky:2013hba} with related discussions.}. Our argument supporting this claim goes as follows. Let us insert a massless quark into the system. In this case the Ward Identity requires that  $\int d^4x\la q(\mathbf{x}), q(\mathbf{0})\ra_{QCD}=0$ in contrast with expression  (\ref{YM}) derived for pure gauge theory. 
   The simplest way to understand this Ward Identity is to represent the topological susceptibility as the second derivative with respect to $\theta$. But physics must be $\theta$ independent in the presence of a massless quark as 
   the $\theta$ parameter can be rotated away by redefinition of the corresponding chiral fermi field. Therefore, $\chi_{QCD}$
   must vanish in the presence of massless quark in the system. How it could happen if physical degree of freedom can only contribute to $\chi_{QCD}$ with the negative sign according to  eq. (\ref{G})? The answer 
 is that this negative conventional dispersive contribution (saturated by the $\eta'$ meson) must cancel with non-dispersive contribution (\ref{YM})  which can not be associated with any physical degrees of freedom. This cancellation is the key element  of the resolution of the  $U(1)_A$ problem~\cite{witten,ven,vendiv}. The explicit computations in this model support this exact cancellation:
   \be
\label{QCD_final}
\la q(\mathbf{x}), q(\mathbf{0})\ra_{QCD} =\frac{\zeta}{NL^2}\left[ \delta(\bold{x})-m_{\eta'}^2\frac{e^{-m_{\eta'}r}}{4\pi r} \right], ~~~ \chi_{QCD}=
\int d^4x \la q(\mathbf{x}), q(\mathbf{0})\ra_{QCD}=0. 
\ee
   The moral is: this ``strange" vacuum energy is very physical and plays a crucial role in resolution of  the celebrated   $U(1)_A$ problem as it  saturates the Ward Identity. In the ``deformed QCD" model this ``strange" energy is resulted from the dynamics of  auxiliary topological field   identified with {\it inflaton}. We treat this contribution  to the vacuum energy (and corresponding fields saturated it) as   physically observable entities as argued above.   
   
 One can also compute a gauge variant  correlation function  
 \be
 \label{K_top}
 \lim_{k\rightarrow 0} \int d^4x e^{ikx} \la \nabla_i a(\mathbf{x}), \nabla_j a(\mathbf{0})\ra \sim\frac{k_i k_j}{k^4}.
 \ee
  The  massless  pole (\ref{K_top}) has precisely the same nature as the pole  in the Veneziano construction  (\ref{K}).  Our comment here is that in spite of the gap in the system, some correlation functions constructed from the {\it inflaton} $a(\mathbf{x}) $ field are still highly sensitive to the IR physics.   Furthermore, while the behaviour (\ref{K_top}) at small $k$ can be considered to be very dangerous as it includes $k^4$ in denominator (which normally attributed to the negative norm states in QFT), the physics described here is perfectly unitary and 
 causal as $a(\mathbf{x}) $ is in fact auxiliary rather than propagating dynamical field as all questions can be formulated and answered even without mentioning the auxiliary topological fields. The behaviour (\ref{K_top}) also hints on possibility of  non-local effects (which indeed are known to be present in this system  \cite{Zhitnitsky:2013hs}). 
 
 \subsection{ ``Strange" energy and the inflaton   field in the expanding universe}\label{odd}
 In this section  we address  very hard question: How does the ``strange" energy (\ref{zeta}) vary when the system couples to the gravity? We can rephrase and simplify the same question  as follows: how does the rate of tunnelling processes change when the system is considered in a time-dependent background? In principle, the strategy  to carry out  the corresponding computations is as follows:\\
 1. find the classical solution in a nontrivial background which generalize  the fractionally charged monopoles 
 reviewed in section \ref{deformedqcd};\\
 2. compute the path integral measure of the corresponding generalized  solutions, similar to the monopole fugacity $\zeta$ from section   \ref{deformedqcd}.
 It includes analysis of zero as well of non-zero modes with corresponding corrections due to the background;\\
 3. compute the interaction between generalized pseudoparticles to present the system in the dual form, similar to eq. (\ref{thetaaction}). The corresponding expression for the effective action should depend now on the parameters of a  background such as the Hubble constant $H$;\\
 4. represent the system in terms of the auxiliary topological fields similar to eq. (\ref{b-action}).
 The corresponding corrections (due to the background) in the coefficients of the action will represent the desired result.
 
 Unfortunately, a solution even of the first step from  this program even in weakly coupled deformed QCD is not known. A resolution of the entire program is not even feasible.  Therefore, the honest answer on the question formulated at the beginning of this section is: the exact answer is unknown. Nevertheless, there are few general arguments which may provide us with  some hints on possible dependence of the ``strange" energy (\ref{zeta}) from  a background which will be parameterized in what follows by the Hubble parameter $H$. In general,  one should expect that for sufficiently weak  background   (which we always assume to be the case) the correction to all observables, including the vacuum energy,  can be represented as the power corrections of $H$, i.e.
 \be
 \label{power}
 E_{\rm vac}(H)=\sum_pc_p H^p, ~~~~ c_0=-\frac{N\zeta}{L}.
 \ee
 
 There are two sets of generic arguments which suggest that $p$ in eq. (\ref{power}) can   only be even, i.e. $p=0, 2, 4...$. The generic arguments, as usual, may have some loopholes....
 The first set of arguments presented in \cite{Kaloper:2002uj}, based on investigation of possible corrections due to the short distance
 physics parametrized by scale $M$.  It has been argued  that corrections should be in form $\sim(H/M)^2$ if the low energy description is {\it local}. 
 At the same time, explicit computations \cite{Easther:2002xe, Easther:2001fz,Easther:2001fi}  in a number of simple  models
 suggest that the corrections could be much larger $\sim(H/M)$, which correspond to $p=1$. Computations \cite{Brandenberger:2002hs,Danielsson:2002qh}
 are also in conflict with generic argument \cite{Kaloper:2002uj}. It is not the goal of the present work to analyze  these conflicting results.
 Rather, we want to point out that sometimes a generic   argument may fail because an assumption may be violated in some hidden way, 
 which   is very difficult to trace. 
 We come back to this point with a similar example which is known to occur in the QCD physics. 
 
 The second set of generic arguments is based on renormalization group analysis, see original papers \cite{Shapiro:1999zt,Shapiro:2000dz,Shapiro:2009dh} and recent review \cite{Sola:2013gha}. The authors of refs \cite{Shapiro:1999zt,Shapiro:2000dz,Shapiro:2009dh,Sola:2013gha} also argue that only even powers may enter eq. (\ref{power}). The arguments  are based on the  {\it locality} and general covariance.  
 We do not wish to analyze in this paper any possible loopholes in general arguments of refs. \cite{Shapiro:1999zt,Shapiro:2000dz,Shapiro:2009dh,Sola:2013gha}. The only comment we want to make is that an assumption of locality might not be so harmless for non-abelian gauge theories such as QCD, in contrast with a simple massive scalar field theory. Indeed, while QCD has a gap in gauge invariant sectors, it nevertheless demonstrates  a high IR sensitivity in gauge variant sectors in terms of a topologically protected massless pole (\ref{K_top}). This pole is not screened by the confinement mechanism and eventually is responsible for the contact term which saturates the ``strange" energy
 (\ref{zeta}) which is the source of {\it inflation} as we shall advocate in the next section. The same IR physics may lead to some non-local effects. In fact, the  weakly coupled deformed QCD model,  reviewed in section \ref{deformedqcd}, 
indeed  show some signs of non-local physics. In particular, in this system one can explicitly demonstrate the presence of the  degenerate states which are classified by a nonlocal operator, while all local expectation values are identically coincide for degenerate states \cite{Zhitnitsky:2013hs}. 
 The presence of such degenerate states in a gapped theory is a typical manifestation of a topologically ordered phase when the system is characterized by a non-local operator. 
 
 Furthermore, with few additional simplifications one can explicitly see how the linear $\sim H$ correction may indeed emerge in the ``deformed QCD" model, see Appendix \ref{H} with some technical details. The main point of the estimates presented in Appendix \ref{H} is that the long ranged auxiliary field  which saturates the ``strange energy" (\ref{zeta}) and which is identified with {\it inflaton} will mix with the background field expressed in terms of the Hubble parameter $H$.  Precisely this mixture leads to the linear correction $\sim H$. A simplified estimate presented in Appendix \ref{H} also demonstrates a deep analogy 
 with the non-local Aharonov -Casher effect mentioned in Introduction as an intuitive picture of the {\it inflation} in our framework. One can explicitly see from Appendix \ref{meaning} how the the {\it non-locality} enters the physics in terms of   inherently {\it non-local}   large gauge transformation operator $\cal{T}$.

 We conclude this subsection with few more  examples which may further  support our main assumption that the correction in eq (\ref{power}) may be order $\sim H$. First example is an explicit computation in the weakly coupled deformed QCD model when it is defined on a finite size manifold $R$ rather than in infinite space-time. Computed correction behaves as $\sim R^{-1}$ \cite{Thomas:2012ib}.
 Our second example is the numerical  lattice   computations in strongly coupled QCD when   the vacuum energy also shows a linear type correction $\sim R^{-1}$  \cite{Holdom:2010ak}.  Our final example is an analysis of the operator product expansion in QCD which 
 suggests that the lowest correction must have dimension $4$ which represents the dimension of $F_{\mu\nu}^2$ operator. 
 This argument is very similar in spirit to general arguments of refs. \cite{Shapiro:1999zt,Shapiro:2000dz,Shapiro:2009dh,Sola:2013gha} mentioned after eq. (\ref{power}).
 It turns out that the corrections could be much larger as a result of the IR sensitive physics when non-locality enters as 
$F_{\mu\nu} \frac{1}{\partial^2}F_{\mu\nu}$ instead of naively  expected behaviour $F_{\mu\nu}^2$ \cite{Gubarev:2000eu}.

  The linear correction  
   can be interpreted in terminology    \cite{Shapiro:1999zt,Shapiro:2000dz,Shapiro:2009dh,Sola:2013gha} as possibility of running cosmological constant at very low $\mu\sim H$ as a result of the IR sensitivity when a non-local physics may emerge as a result of non-locality  of the operator $\cal{T}$, see Appendix \ref{meaning} with some details.  This renormalization is obviously a non-perturbative in nature, as all effects discussed in this work, including the 
    ``strange" energy,   can not be seen at any level in perturbation theory as $\zeta\sim \exp(-1/g^2)$ as they are originated from the deep IR physics.

\section{  Scaled up $\qcd$  and  Inflationary de Sitter Phase}\label{inflation}

This is the main section of the present work. 
Based on the arguments  presented in previous section \ref{odd}  and Appendix \ref{QCD}  we assume in what follows that the first   non-vanishing correction to the ``strange" vacuum energy scales as $\sim H$  when the scale up version  $\qcd$  is defined in the expanding background characterized by the Hubble parameter $H$. In other words, the expression for the vacuum energy 
 in context  Friedmann-Lema\^itre-Robertson-Walker (FLRW) universe assumes the following form
  \be
 \label{FLRW}
  E_{\mathrm{FLRW}}(H)\sim \left[\Lbar^4+ H\Lbar^3+ {\cal{O}}(H^2)\right], 
  \ee
  similar to our analysis of the weakly coupled deformed QCD model (\ref{power}).
There are two  crucial points here:

1. The corrections start with a linear term $\sim H$.  The source of this linear term as argued  above is the inherent non-locality of the large gauge transformation operator $\cal{T}$ which itself is the key element in the mechanism of generating the ``strange" energy.  We shall see in a moment that this term $\sim H\Lbar^3$ will drag the universe into de Sitter state.
  
  2. The ``strange energy" (\ref{FLRW}) vanishes in deconfined $\qcd$ phase above the phase transition. The corresponding inflaton field $\phi(x)$, generating (\ref{FLRW}), and  replacing the $a(\mathbf{x})$ field from section \ref{BF-section},  does not exist in the deconfined $\qcd$ phase, see Appendix \ref{QCD} with specific technical details. This should be contrasted with conventional inflationary models when field $\Phi$ and corresponding potential $V(\Phi)$ always existed in the system.  

\subsection{On interpretation of the ``strange"  energy}\label{interpretation}
Before we start with the computations we want to elaborate on the physical meaning of $E_{\mathrm{FLRW}}(H)$. We interpret this energy as the energy which is generated due to the tunnelling processes when transitions are happening all the time between topologically different but physically identical states as explained in section \ref{BF-section}.
 When the system is placed into the FLRW  background   the corresponding {\it rate} of transitions changes as a result of Hubble expansion. This variation of the rate    is reflected by eq. (\ref{FLRW}) in form of $H$-dependent corrections. The source of this ``strange" energy can be represented in terms of the {\it inflaton} field $\phi(x)$ with a number of unusual features,  as discussed in section \ref{BF-section}. However, we should emphasize that 
 the energy  (\ref{FLRW}) 
 is very different in nature  from conventional vacuum energy determined by the vacuum expectation value $\la\Phi\ra$ and its potential
 $V(\la\Phi\ra)$, similar to the Higgs model. In particular,   the energy  (\ref{FLRW}) 
 can not be formulated in terms of a dynamical field $\Phi$ with canonical  kinetic term and local potential $V(\Phi)$ as  explained in section 
 \ref{BF-section} using weakly coupled deformed QCD model as a theoretically treatable example. This feature is a simple reflection of the fact that the physics of tunnelling processes and the corresponding generated energy can not be described in terms of  a local dynamical field $\Phi$, as the tunnelling  between topologically distinct sectors is fundamentally non-local phenomenon. Furthermore, the energy (\ref{FLRW}) vanishes above $\qcd$ phase transition   in deconfined phase as this structure emerges  only as a result of confinement. In other words, our auxiliary fields $\left[\phi(x), b(x)\right]$    do not exist in deconfined phase. This is again in a huge contrast with conventional inflationary scenarios when 
 $V(\Phi)$ always existed, before, and  after the inflation.

 Our second comment before we start our computations is the formulation of the prescription that 
  the relevant energy which enters the Einstein equations is in fact 
 the difference $\Delta E\equiv E -E_{\mathrm{Mink}}$ between the energies  of a system in a non-trivial background  and Minkowski space-time geometry,  similar to the well known Casimir effect when the observed   energy is  a difference 
  between the energy computed for a system with conducting boundaries   and infinite Minkowski space.  In this framework it is quite natural to define the ``renormalized vacuum energy'' to be zero in Minkowski vacuum wherein the Einstein equations are automatically satisfied as the Ricci tensor identically vanishes.   
  
  In the present context  such a definition $\Delta E\equiv (E_{FLRW} -E_{\mathrm{Mink}})$ for the vacuum energy for the first time was advocated   in 1967   by Zeldovich~\cite{Zeldovich:1967gd} who argued that  $\rho_{\text{vac}}=\Delta E \sim Gm_p^6 $ with $m_p$ being the proton's mass. Later on such  definition for the relevant energy $\Delta E\equiv (E_{FLRW} -E_{\mathrm{Mink}})$ which  enters the Einstein equations has been advocated from   different perspectives in a number of papers,  see e.g.  relatively recent works~\cite{Bjorken:2001pe, Schutzhold:2002pr, Klinkhamer:2007pe, Klinkhamer:2009nn, Thomas:2009uh,Maggiore:2010wr}, see also review article \cite{Sola:2013gha} with 
  a background on the subject and large number of references. This prescription is consistent with
the renormalization group approach   advocated in  \cite{Shapiro:1999zt,Shapiro:2000dz,Shapiro:2009dh,Sola:2013gha}. In  fact, it is direct consequence of the  renormalization group approach when we   fix a physical parameter at one point of normalization to predict 
its value at a different normalization point. In context of eq. (\ref{FLRW}) it implies that   the vacuum energy which enters the  Einstein equations is  $\Delta E\equiv (E_{FLRW} -E_{\mathrm{Mink}})$ at normalization point $\mu\sim H$.  As we already mentioned, this prescription is consistent with the  Einstein equations 
when the vacuum energy  approaches zero,  $\Delta E\rightarrow 0$  for Minkowski space-time geometry which itself may be considered as a limiting case with $H\rightarrow 0$. 

Our final comment before we start the computations goes as follows. The energy (\ref{FLRW}) can be interpreted as a running cosmological constant within the renormalization group approach   advocated in  \cite{Shapiro:1999zt,Shapiro:2000dz,Shapiro:2009dh,Sola:2013gha} with the only difference that odd powers of $H$ are also included into the series as a result of the IR sensitivity and non-locality  as discussed  in section \ref{odd} and Appendix \ref{H}. 
This linear correction  
   can be interpreted in terminology    \cite{Shapiro:1999zt,Shapiro:2000dz,Shapiro:2009dh,Sola:2013gha} as possibility of running cosmological constant at very low $\mu\sim H$. This running is  originated from non-perturbative and non-local physics  and can not be seen at any finite level in perturbation theory, as entire ``strange" energy can  not  be generated   in perturbation theory. Nevertheless, all terms in expansion (\ref{FLRW}) are finite and uniquely defined, similar to our discussions in a simplified model in section \ref{BF-section} where all computations are under complete theoretical control.   Furthermore, this energy is not generated during the deconfined phase, as it starts to emerge only in confined $\qcd$ phase. 

 \subsection{Inflationary de Sitter phase}

With these preliminary remarks on $\qcd$ and its relation to cosmology, we can now write down  the Friedman equation    as follows
\be
 \label{friedman-infl}
 H^2&=& \frac{8\pi G}{3}\left(  \rho_{\mathrm{Inf}}+\rho_R\right)=
 \frac{8\pi G}{3}\left( \overline{\alpha} H\Lbar^3+\rho_R\right), \nonumber\\
  \rho_{\mathrm{Inf}} &=& \overline{\alpha}H\Lbar^3, ~~~   \rho_R= \frac{\pi^2}{30} N(T)T^4, ~~~ N(T)=N_b(T)+\frac{7}{8}N_f(T)
 \ee
where we introduce notations for   the corresponding  energy density   $\Delta E= \overline{\alpha}H\Lbar^3$  
with $\overline{\alpha}$ being a dimensionless parameter of order one and we neglected higher order correction $ {\cal{O}}(H^2)$.
We also introduce notation $ \rho_{\mathrm{Inf}}= \Delta E$ to emphasize that   this term will drive the universe to the de Sitter inflationary phase as we shall see in a moment. 

The inflation in this framework starts from thermal equilibrium state with $N(T)$ massless degrees of freedom (at time of inflation)    which will be eventually responsible for reheating to be discussed in section \ref{reheating}. If we identify 
these massless degrees of freedom  with standard model (SM) particles than $N(T)\sim 10^2$. We note that SM fields are  indeed almost massless at high  temperature when inflation starts, but they become massive, except photon, at present temperatures. It is also possible that some other fields, beyond SM particles are massless at such high temperature, but we do not speculate  on this point in the present work. As we shall see in a moment, the inflation in this framework starts  
long after the $\qcd$ phase transition such that $N(T)$ in eq. (\ref{friedman-infl}) does not include any $\qcd$ physical states  as they are heavy at that time.  It is important  to emphasize that the corresponding energy $ \overline{\alpha}H\Lbar^3$ emerges soon after the $\qcd$ phase transition. 
However,  the energy  $\overline{\alpha}H\Lbar^3$ starts to compete with $\rho_R$ at much later times, when $\qcd$ is in a deep confined regime.  

The radiation component in eq. (\ref{friedman-infl}) scales as $\rho_R\sim {\rm a}^{-4}$ such that $\rho_{\mathrm{Inf}}$ starts to dominate the universe at some point when $H$ approaches the constant value $H_0$, see estimate below.  
 This state of evolution of the universe is a starting point of the  inflationary regime. To quantify the analysis   we  shall introduce  a subscript $0$ in $\rho_{R0}$ for the value when $\rho_{R0}= \rho_{\mathrm{Inf},0}= 1/2 \rho_c$ and ${\rm a}_0=1$.
 In different words,  subscript $0$ shows the moment in evolution of the universe when energy density related to inflation  becomes the dominating component exceeding the radiation component.

The Hubble parameter $H_0$ and  the temperature $T_0$ when the inflation effectively starts in this model can be 
 estimated as follows
 \be
 \label{T}
 H_0\sim \frac{8\pi G}{3} ( \overline{\alpha} \Lbar^3 ), ~~
   \rho_{R0}\simeq \rho_{\mathrm{Inf},0}\simeq \frac{1}{2}\rho_c  ~~ \Rightarrow ~~ T_0\simeq  \Lbar\sqrt{ \overline{\alpha}\frac{\Lbar}{M_{PL}}} \left(\frac{40}{\pi N}\right)^{1/4},
 \ee
where $M_{PL}$ is defined as usual, $M_{PL}=1/\sqrt{G}$. Assuming that $T_0$ is much higher than Electro-Weak scale, $T_0\gg M_{EW}$ one can estimate 
a lower bound for the $\qcd$ related  physics determined by a new scale $\Lbar$ 
  \be
 \label{bound}
 \Lbar\gg \sqrt[3]{M_{EW}^2M_{PL}}\sim 10^8~ {\mathrm{GeV}}.
 \ee
As anticipated, 
\be
\frac{T_0}{\Lbar}\sim\sqrt{ \overline{\alpha}\frac{\Lbar}{M_{PL}}} \left(\frac{40}{\pi N}\right)^{1/4} \ll 1,
\ee
 and therefore the physical massive $\qcd$ degrees of freedom indeed do not participate in the thermo- dynamical equilibrium when  inflation effectively starts in this model 
at $T_0$ and do not contribute to $N(T)$ as stated above.

One can solve  the Friedman equation (\ref{friedman-infl})  with the following result
  \be
 \label{infl_s}
 H=\frac{4\pi G}{3}\overline{\alpha}\Lbar^3 +\sqrt{\left(\frac{4\pi G}{3}\overline{\alpha}\Lbar^3\right)^2+ \frac{8\pi G\rho_{R0} }{3{\rm a}^4} } , 
 \ee
To analyze the solution of this equation it is convenient to  define 
  a characteristic scale $\bar{\rm a} _{\star}$ when two terms under the square root in eq. (\ref{infl_s}) become equal
\be
 \label{infl_a}
 \bar{{\rm a}}_{\star}^4= \frac{3}{2\pi G} \frac{\rho_{R0}}{\left(\overline{\alpha}\Lbar^3\right)^2}.
 \ee
In terms of these parameters the behaviour of the energy density $  \rho_{\mathrm{Inf}}$ related to the inflation  can be conveniently represented as follows 
  \be
 \label{infl-de-sitter}
 \rho_{\mathrm{Inf}} =\overline{\alpha}H\Lbar^3= \frac{4\pi G}{3}\overline{\alpha}^2\Lbar^6\left[1+\sqrt{1+\left(\frac{\bar{{\rm a}}_{\star}}{{\rm a}}\right)^4}\right].
 \ee
 One can explicitly see from this solution that for ${\rm a}\ll \bar {\rm a}_{\star}$ the radiation  component dominates in eq. (\ref{friedman-infl}),
 while for ${\rm a}\gg \bar {\rm a}_{\star}$ the inflation  component dominates with the following asymptotic behaviour 
 \be
 \label{infl-de-sitter1}
 \rho_{\mathrm{Inf}}= \frac{8\pi G}{3}\overline{\alpha}^2\Lbar^6\left[1+\frac{1}{4}\left(\frac{\bar{{\rm a}}_{\star}}{{\rm a}}\right)^4\right],  ~~~~~~ H = \frac{8\pi G}{3}\overline{\alpha}\Lbar^3\left[1+\frac{1}{4}\left(\frac{\bar{{\rm a}}_{\star}}{{\rm a}}\right)^4\right], ~~~ {\rm a}\gg \bar{{\rm a}}_{\star}. 
 \ee
 As stated previously, the Hubble parameter is approaching the constant value $H_0$ at asymptotically large ${\rm a}\gg \bar{{\rm a}}_{\star}$. In different words, the evolution of the universe in this model approaches a  de-Sitter state
  at asymptotically large ${\rm a}\rightarrow\infty$ 
  as claimed above.  The radiation component can be also easily computed in this framework. Its asymptotical behaviour is given by
   \be
 \label{rad-de-sitter1}
 \rho_{R}= \frac{2\pi G}{3}\overline{\alpha}^2\Lbar^6 \cdot \left(\frac{\bar{{\rm a}}_{\star}}{{\rm a}}\right)^4,  ~~~ {\rm a}\gg \bar{{\rm a}}_{\star}. 
 \ee
  As expected the radiation becomes a subdominant component for large ${\rm a}\gg \bar{{\rm a}}_{\star}$.
  As an explicit expression for $H$ is known   one can  explicitly  compute the equation of state (EoS) for inflationary phase in this system. To simplify formula we only consider the asymptotical behaviour at large ${\rm a}\gg \bar{{\rm a}}_{\star}$. In this case one can differentiate  eq. (\ref{infl-de-sitter1}) and substitute to a general equation for 
  $\dot{H}=-4\pi G (\rho+p)$ to arrive to the following expression
    \be
  (\rho+p)= \frac{2}{3}\cdot\overline{\alpha}\Lbar^3 H\cdot \left(\frac{\bar{{\rm a}}_{\star}}{{\rm a}}\right)^4, ~~~~ {\rm a}\gg \bar{{\rm a}}_{\star}.
 \ee
  One can   represent   this  EoS for the inflationary (almost) de Sitter behaviour in the following conventional  form
  \be
 \label{infl-EoS}
  \omega \equiv \frac{p}{\rho}\simeq -1+\frac{2}{3}\left(\frac{ \bar{{\rm a}}_{\star}}{{\rm a}}\right)^4, ~~~ {\rm a}\gg  \bar{{\rm a}}_{\star}
 \ee
 such that the EoS will   approach $-1$ from above, and the universe is dragged into a de- Sitter state
 at asymptotically large ${\rm a}$. In fact, the scale factor growth exponentially fast
 already in close vicinity of  ${\rm a}> \bar{{\rm a}}_{\star}$ as eq (\ref{infl-EoS}) suggests. Therefore, 
 with very good accuracy, one can use the following expression for scale factor   ${\rm a}(t)$  for all ${\rm a}> \bar {\rm a}_{\star}$
 (though it is formally valid only
 for ${\rm a}\gg \bar{{\rm a}}_{\star}$),
 \be
 \label{a2}
   {\rm a}(t)\sim \exp (H_{\infty}t)   , ~~~~ H_{\infty}=  \frac{8\pi G}{3}\overline{\alpha}\Lbar^3, ~~~~ \omega=-1, 
   ~~~ {\rm a}> \bar{{\rm a}}_{\star},
 \ee
 where $H_{\infty}$ is determined by   eq. (\ref{infl-de-sitter1}) at asymptotically large ${\rm a}$.  In other words, it takes only a single e-fold (single Hubble time $\sim H_{\infty}^{-1}$ ) in evolution of the universe when  the de-Sitter behaviour (\ref{a2}) effectively  becomes fully operational, and formula (\ref{a2}) can be used during entire inflationary regime as Hubble constant $H$ indeed stays  almost constant during the inflation.  
 
 We conclude this section with two  comments. First, equation similar to 
 eq. (\ref{FLRW}) was previously postulated in \cite{dyn, 4d}  (admittedly, with very little  understanding what is behind this formula\footnote{\label{DE}In particular, the fact that the system does not violate gauge invariance, unitarity, causality was demonstrated   in a follow up paper \cite{Zhitnitsky:2010ji}, see also \cite{ohta} with related discussions.  In present formulation in terms of the auxiliary fields from section \ref{BF-section} these features are trivially satisfied  as the entire system can be reformulated in terms of auxiliary topological non-propagating fields when the questions on unitarity and  causality  do not even emerge. The question on possibility of linear -like corrections due to the background field were later addressed  in the holographic QCD model in \cite{Zhitnitsky:2011aa} and  computed in deformed QCD model in \cite{Thomas:2012ib}. Such linear-like corrections  were also supported by the lattice studies \cite{Holdom:2010ak}. Finally, 
 as recently discussed in \cite{Zhitnitsky:2013hs} the system may demonstrate some non-local features. 
 This non-locality and IR sensitivity  may falsify  the main assumption leading to the conventional $H^2$ prediction,   as discussed in section \ref{odd} and Appendix \ref{H}.})  to describe the dark energy as a result of the QCD dynamics.  Most importantly, this postulate  has been (successfully) confronted with observations, see \cite{Cai:2010uf,Sheykhi:2011xz, Sheykhi:2011nb,RozasFernandez:2011je,Cai:2012fq,Alcaniz:2012mh,Feng:2012gr,Malekjani:2012wc,Feng:2012wx} and many references therein, where it has been claimed that 
 this model is consistent with all presently available data\footnote{\label{c_s}A short warning signal is as follows: the authors of  some   papers mentioned above   use the auxiliary quantum fields as the classical fields which satisfy the classical equations of motion. This is obviously a wrong procedure. In particular, a computation of derivative $(\partial p/\partial \rho)$  and identification it with the speed of sound $c_s^2$ makes really no sense as there is no any propagation with such speed because there are no any physical propagating degrees of freedom in the system. It is quite obvious that one can not interpret $c_s^2<0$ in such a computation as instability of a system.}. 

Our second comment is as follows.  In the analysis presented in this section we completely ignored the interaction with other fields.   If no other light fields interacting with $[\phi(x), b(x)]$ are present in the system, the regime (\ref{a2}) would be the final destination of our universe.   However the interaction of auxiliary $[\phi(x), b(x)]$ fields with light  particles does exist  in this system, and, in fact, the end of inflation   is triggered precisely by this interaction.  As we shall see  below  the corresponding relevant coupling   is  unambiguously  fixed by the  well known triangle anomaly and transformation properties of the path integral under the chiral transformations. However, the ``theory of reheating" is still to be developed for this framework as it is fundamentally different in nature from conventional picture when  a dynamical inflaton $\Phi$   transfers
its energy to light particles. Therefore,  we can not borrow the  technique  \cite{linde,mukhanov} which is well developed  for the  conventional inflaton models.  Nevertheless, we opted to sketch  some thoughts on this matter with a hope that it may help to  develop  the theory of reheating within $\qcd$-{\it  inflationary} proposal in future.
 
 \section{Few thoughts on Reheating   }\label{reheating}

It is well known that  for  the inflation to   end, one should couple the relevant fields responsible for inflation (in our case this role is played by $ [\phi(x), b(x)]$ fields) with light degrees of freedom of the standard model  such that the energy generated  during inflation can be released by producing   particles and radiation.  This is  so-called the reheating period. To simplify things we assume that  $\qcd$  has one quark in fundamental representation which interacts with the $E\&W$ gauge bosons  precisely in the same way as conventional QCD quarks do. In this case all couplings and algebraic structures of the interacting terms are unambiguously fixed. The conventional interactions $A_{\mu}J^{\mu}$ of $\qcd$ quarks with $E\&W$ gauge bosons  is no relevance for our purposes as the    fields responsible for inflation are  in fact  auxiliary topological  fields $ [\phi(x), b(x)] $ which   interact with SM particles only as a result of anomalous coupling with a background. Therefore, in what follows we only consider the interaction of the SM particles with  auxiliary $ [\phi(x), b(x)]$ fields responsible for inflation.
 
 The relevant for this paper  interaction of the $  b(x)$ field with SM particles occurs  as a result of anomaly. 
 To simplify things further we consider the interaction of the $  b(x)$ field with $E\&M$ photons only. A coupling  of the $b(x) $ with other gauge bosons can be unambiguously  reconstructed using WZNW Lagrangian \cite{wznw}, but we keep a single  $E\&M$ field  $F_{\mu\nu}$ to simplify the notations and outline the idea on possible  reheating mechanism. 
In our context   the corresponding coupling has the following form 
  \be\label{coup}
{\cal L}_{b\gamma\gamma} = \frac{\alpha (H_0)}{8\pi} N  Q^2 \left[ \theta- b(x)\right] \cdot F_{\mu\nu} \tilde F^{\mu\nu} \, ,
\ee
where $\alpha(H_0)$ is the fine-structure constant measured at moment $H_0$, i.e. during the period of inflation,   $Q$ is the electric charge of the $\qcd$ quark, and  $F_{\mu\nu}$ is the usual electromagnetic field strength.   The coupling (\ref{coup}) is unambiguously fixed because   the auxiliary $b(x)$ field always accompanies the $\theta$ parameter in a specific combination $\left( \theta- b(x)\right)$  as explained in Appendix \ref{QCD}.  The coupling (\ref{coup}) describes the anomalous  interaction of the topological auxiliary $b(x)$  field with $E\&M$ photons.
 We  assume $\theta=0$ in eq. (\ref{coup}) once   coupling with $b(x)$ field is reconstructed.        
 
  One should remark   here that  a similar coupling of the photons with the axion $\theta(x)$ field in context of inflationary cosmology was considered long ago \cite{Freese:1990rb} with many followup proposals. It  has been also known that this interaction  
  leads to   instability with respect to particle  production and radiation.  Therefore, the interaction (\ref{coup}) potentially may serve as a source of reheating.
  The crucial difference of  the present studies with ref. \cite{Freese:1990rb} is that our field $b(x)$ is not a dynamical field, similar to a physical propagating axion field considered in \cite{Freese:1990rb}. Rather, it is an auxiliary topological $b(x)$ field which does not propagate and has no  kinetic term. It other words, the instability with respect to radiation may occur in our system  not due to the fluctuations of a dynamical (pseudo)scalar field.  Rather,   the corresponding radiation  might be generated as a result of fluctuations of the auxiliary $b(x)$ field. 
  Therefore, the underlying dynamics of the fluctuations eventually leading to the radiation   (reheating epoch) is fundamentally different from conventional radiation by  a propagating 
  (pseudo)scalar   axion field. 
  
  The corresponding ``theory of reheating" within $\qcd$-{\it  inflationary} proposal is yet to be developed.   
  In this framework  the  $b(x)$-field  should be treated as a coherent field representing the rate of tunnelling events in the system. It  varies and  fluctuates as a consequence  of expansion, rather than a result of the presence of a kinetic term.  As a result of these fluctuations in time dependent background $b(x)$ field radiates real physical particles in expanding universe.  This radiation occurs  in spite  of the fact that $b(x)$ itself is not a dynamical field. This is precisely the way how the energy (generated due to the tunnelling processes and  expressed in terms  of auxiliary  $ [\phi(x), b(x)]$ fields) in principle can be transferred   to   the SM particles. 
  In weakly coupled deformed QCD model the corresponding computational procedure is outlined as steps 1-4
  in section \ref{odd}. These computations, in principle, should predict the dynamics of the fluctuating  auxiliary topological  fields $ [\phi(x), b(x)] $ in expanding universe.  
  Eventually this process of the energy transfer should be  responsible for  the termination of the inflationary epoch.  
  
  As we already mentioned we do not have a developed machinery to carry out such computations along the line outlined above. However, we can make few simple estimates and provide some analogies with a physical system which is known to exist in nature and realized in heavy ion collisions, see below.
    
  It is clear that the relevant scale which enters the problem is $H$ during the inflation time, rather than $\Lbar$ scale itself.
  Indeed,  the expectation value for the $b(x)$ field obviously vanishes in Minkowski space time $\la b(x)\ra=0$. Furthermore,  no radiation of physical photons  may occur in Minkowski vacuum even though the $ [\phi(x), b(x)]$ fields do fluctuate to saturate
  the $H$-independent term in expansion (\ref{FLRW}). In other words, all  effects which lead to the radiation  must be proportional to small corrections $\sim H$ exclusively due to expansion, similar to (\ref{FLRW}). The same conclusion
  also follows from the observation that for a constant $b$ in eq. (\ref{coup}) the Lagrangian  represents a total derivative and can not lead to any radiation, such that physical effects must be proportional to $\dot{b}$. 
  In this framework the number of e-foldings in the $\qcd$-{\it  inflation} is determined by the time  $\tau_{inst}$ when the instability due to the  radiation is fully developed.
  This is exactly the time scale when  the entire energy (\ref{infl-de-sitter}),(\ref{infl-de-sitter1}) generated during the inflation is transferred to SM light fields.  
  
  To estimate the time scale  $\tau_{\rm inst}$ we note that  $\tau_{\rm inst}^{-1}\sim H$ must be proportional to $H$ as the only relevant scale of the problem as explained above. Furthermore  the effect must be proportional to coupling constant with some power $k$, i.e.$\tau_{\rm inst}^{-1}\sim \alpha^k$. In fact, we expect that $k=2$  as a similar phenomenon  
  with identically the same interaction (\ref{coup}) 
  has been actually discussed in the literature in context of heavy ion physics, see details in Appendix \ref{ions}.
  The role of the auxiliary $b(x)$ field from eq. (\ref{coup}) is played by the so-called axial chemical potential, which is also an auxiliary field in the  heavy ion physics (\ref{mu5}). 
  Combine all factors together we arrive to the following estimate  
    \be
    \label{e-folds}
  \tau_{\rm inst}^{-1}\sim  {H\alpha_s^2}, ~~~~~\Longrightarrow ~~~~~~ \tau_{\rm inst}\sim  \frac{1}{H\alpha_s^2}
   ~~~~~\Longrightarrow ~~~~~~N_{\text{Inf}}\sim \frac{1}{\alpha_s^2},
  \ee   
  where number of e-folds $N_{\text{Inf}}$ is, by definition, the coefficient in front of $H^{-1}$ in the expression for $\tau_{\rm inst}$. 
    In Appendix \ref{ions} we discuss an analogy with very similar problem of the helical instability studied in heavy ion physics. The relevant point,   for our present estimates (\ref{e-folds}),  is that a similar instability also develops in strongly coupled gauge theory as the energy can be transferred not only to photons but to $\bar{q}q$ pairs as well. In this case, the instability develops much faster
  as it will be determined by the strong coupling $\alpha_s$ which we expect to enter eq. (\ref{e-folds}). In any event, at  the inflationary scale the strong coupling constant $\alpha_s$ and the weak   coupling constant  $\alpha_w$ do not differer much and numerically, up to factor two, very close to each other. Therefore, we do not distinguish between them in our very crude dimensional estimate (\ref{e-folds}). The important point is that
   the estimate (\ref{e-folds}) shows that  $N_{\text{Inf}}$ could   easily achieve the required number of e-folds $N_{\text{Inf}}\geq 70$
 as the strong coupling $\alpha_s\sim 0.1$ is already small at the inflation scale. 
 
 We want to repeat once again: the estimate (\ref{e-folds}) must be taken with a grain of salt as it is essentially based on dimensional analysis, while a solid computation machinery is yet to be developed   as outlined above. We   present this estimate exclusively with a demonstration purpose to emphasize  that the number of e-folds $N_{\text{Inf}}\geq 70$ might be related to the gauge dynamics and expressed in terms of 
 a small  gauge coupling constant   within  $\qcd$-{\it  inflation} framework, rather than it is related to 
the  properties of the classical inflaton  potential $V(\Phi)$ with the corresponding slow-roll requirements.   What is more important is that we anticipate that this number should be expressed (eventually) in terms of the gauge coupling constants we know and love.

  As we emphasized above: the theory of reheating in  $\qcd$-{\it  inflation} is yet to be developed.  Therefore, we do not know 
  answers on many relevant questions\footnote{For example, it is obvious that the $\rho_{\rm Inf}$ can not stay the same while its energy flows to radiation as a result of interaction, while formula (\ref{friedman-infl}) suggests its (almost) constant value expressed in terms of $\Lbar$. It is clear that all formulae presented above are written by ignoring the interaction and assuming an instant ``equilibribration" when the gravity immediately fills up the $\rho_{\rm Inf}$  portion of energy which was just used  as a result of radiation. In reality it must be clear  that it takes some time to fill this energy, which is obviously very long process as the corresponding energy transfer  is proportional to the gravitational constant, while removing this energy is a much faster process as it is proportional to a gauge coupling. Answering this and many other questions would eventually predict the fate of the universe. The corresponding analysis is well beyond the scope of the present paper, as it requires the understanding of many problems  such as  reheating, back reaction and many other related questions within this framework. }.
  Nevertheless, we anticipate that all small parameters which are normally required  for successful inflation will be (eventually) expressed  in our framework in terms of a small gauge coupling constant during the  reheating time, because  
  precisely this interaction modifies   the EoS (\ref{infl-EoS}) by producing small corrections $\sim \alpha_s^2$, similar to (\ref{e-folds}).
  Density perturbations in this framework is generated by the auxiliary topological fields $ [\phi(x), b(x)]$  which is responsible for the (almost) de-Sitter behaviour (\ref{a2}).   The standard prediction for all inflationary models (including our framework) is that the fluctuations are (almost) scale invariant as a consequence of the de-Sitter symmetries during the inflation phase \cite{mukhanov}. Therefore, we have not much new to say regarding this standard and very generic prediction of the inflationary idea. While prediction on scale invariance of perturbations is identical to conventional inflationary models, a computational scheme for the size of the perturbations in our framework is very different from the standard procedure. The same comment  also applies to estimation of the spectral index $n_S$ and its deviation from unity, which we expect to  be expressed in terms of a gauge  coupling constant, similar to eq. (\ref{e-folds}), i.e.  $|n_S-1|\sim \alpha_s^2$, see recent ref. \cite{Zhitnitsky:2014aja} with some computations along this line.
  
  All these hard problems are reduced in our framework to study of the equation of state (\ref{infl-EoS}) at the end of inflationary phase when helical instability develops and the interaction plays a key role. Therefore, the corresponding corrections should be proportional to the gauge coupling constant  similar to our dimensional  analysis  of $N_{\text{Inf}}$.  We anticipate that the relevant technique to study these hard questions will be similar in spirit to the technique employed  in study of the helical instability in heavy ion physics and reviewed in Appendix \ref{ions}: in both cases the instability leads to the decreasing  an auxiliary  $\dot{b}(x)$ field which was the original source of instability. In heavy ion physics the corresponding auxiliary field is identified with the axial chemical potential (\ref{mu5}), which is indeed is getting reduced as a result of the instability. In our cosmological context such flow  of energy   implies that the fate of instability is to reduce the inflationary Hubble constant  (\ref{a2}). The corresponding inflationary energy which is proportional to $H$ will be transferred to the light particles, which is precisely the destiny and fate  of the reheating epoch.

   \section{Conclusion. Future Directions.}\label{conclusion} 
   In the present work we advocate an idea that the  {\it inflaton} field is not a fundamental local field. Instead, the role of the {\it inflaton}
   plays an auxiliary   topological field which effectively describes the dynamics of topological sectors in the  gauge theory when it is considered in the expanding universe.  The corresponding energy in this framework has fundamentally different nature than conventional energy  when a theory is formulated  in terms of a fundamental dynamical  field $\Phi$, for example in the Higgs model. In particular, it can not be expressed in terms of any propagating physical degrees of freedom as the corresponding energy  has a non-dispersive nature\footnote{\label{lattice} The corresponding physics is well understood in QCD. In particular, the  non-dispersive  contribution to the topological susceptibility  with the ``wrong sign" , and corresponding the $\theta$ dependent portion to the energy    are well confirmed by the lattice studies, see e.g.  \cite{Zhitnitsky:2013hs} with large number of references on the original lattice studies. It is also well known that the topological susceptibility (and the energy associated with it) vanishes in deconfined phase. We use a simplified gauge theory, the weakly coupled deformed QCD, reviewed in sections \ref{deformedqcd}, \ref{BF-section} to explain all these ``strange" features  using the auxiliary $ [\phi(x), b(x)]$ fields. It provides us with some simple intuitive picture of the system which is difficult to explain using  the original numerical lattice QCD results.}.    
   Similar auxiliary non-propagating topological fields are known to play an important role in many condensed matter systems realized in nature.   The energy in our system is generated due to the tunnelling processes describing the transitions between   topologically different but physically equivalent winding states. The {\it inflaton} field which effectively describes these transitions in expanding background emerges after the confinement -deconfinement phase transition. This field ceases to exist in deconfined phase, in contrast with all conventional inflationary scenarios  when local $\Phi$ field and its potential $V(\Phi)$  always exist in the system.  This topological field does not have  kinetic term, and it does not propagate as it is an auxiliary field. These features  in fact may  have some profound observational consequences as we shall argue below. \\
   
    {\bf  \underline{Assumptions.}} Our construction is  based on   three basic assumptions:\\
 1. We assume there existence of a scaled up version  of QCD which is coined in this paper as $\qcd$. It is not really a  very new idea as similar construction (though in a different context)
  has been suggested  long time ago and it is known as   technicolor, see recent review  article~\cite{Sannino:2009za}. We do not discuss any connections  with  technicolor models   in  present paper.
 However,  in principle the corresponding studies might be  worthwhile  to explore as the $\Lbar$ scale (\ref{bound}) could be quite appropriate for  these purposes. The only constraints on  $\qcd$ are: it must be asymptotically free gauge theory to satisfy UV completion requirement, and also $\Lbar \gg 10^{8}$ GeV  to avoid interference with $E\&W$ physics.\\
  2. We adopt the paradigm that the relevant definition of  the energy   which enters the Einstein equations 
 is $\Delta E\equiv (E -E_{\mathrm{Mink}})$, similar to the Casimir effect. This is absolutely consistent procedure for  formulating of a QFT in a curved background as discussed in section \ref{interpretation}.
 This element in our analysis is also not very new, and in fact  in the present context  such a definition for the vacuum energy for the first time was advocated   in 1967   by Zeldovich~\cite{Zeldovich:1967gd},
  see  \cite{Sola:2013gha}
for review. \\
  3. A novel element which was not widely discussed previously in the literature is an assumption that  the 
  ``strange" vacuum energy (\ref{FLRW}) receives the linear corrections $\sim H$  in apparent contradiction with conventional arguments that the corrections must be quadratic $\sim H^2$, see section \ref{odd} with details on    pros and cons of each argument. 
An explicit computation which could resolve this issue even in a weakly coupled toy model is  hard to carry out, see  steps 1-4 in section \ref{odd}. 
 Similar in spirit the non-local features are  known  to be present  in many gapped topologically ordered condensed matter systems realized in nature.  This non-locality may falsify the main assumption leading to $H^2$ prediction as argued in section \ref{odd} and Appendix \ref{H}. \\

  {\bf  \underline{Basic result.}} With these  three  assumptions   just formulated, we have argued that the universe  had a period of inflationary  (almost) de Sitter phase 
  characterized by behaviour (\ref{a2}). We also argued that  the regime (\ref{a2}) would be the final destination of our universe
  if    interaction with   SM fields is switched off. When the coupling is switched back on, the end of inflation is triggered
precisely by this  interaction which itself is unambiguously fixed by triangle anomaly. We also presented an order of magnitude estimates based on dimensional arguments for number of e-folds (\ref{e-folds}).  
 \\
    
   {\bf   \underline{Other profound  consequences    of the framework.}}  
 The origin for the de Sitter behaviour (\ref{a2}) is  obviously very different from conventional  inflationary scenario normally formulated in terms of a scalar dynamical field $\Phi$, see  recent review papers  \cite{Linde:2014nna,Brandenberger:2012uj} with   opposite views on   inflationary  cosmology.  For example, as is known, the initial value of the inflaton field $\Phi_{\rm in}$ (in conventional scenario) must be larger than Plank scale to provide a sufficient number of e-folds $ N_{\rm Inf}\sim (\Phi_{\rm in}/M_{\rm PL})^2$. A similar  constraint is also required to support  a  slow-roll condition. 
  In our framework, the relevant $\qcd$ scale never becomes above the Planck mass, while the number of $e$-folds is determined by the gauge coupling constant (\ref{e-folds}).
  Still, both mechanisms, the $\qcd$- inflation and conventional 
approach  \cite{Linde:2014nna} eventually  lead to the same de Sitter behaviour (\ref{a2}).
It would be very interesting to analyze and study the possible observational differences between these two fundamentally distinct  frameworks. 

    {\bf   \underline{Related effects}}  
We conclude this work with mentioning two related phenomena which are similar in spirit, but characterized by the drastically different scales. First, as we already mentioned the energy described by a formula similar to eq. (\ref{FLRW}) (which  eventually leads to the de Sitter behaviour (\ref{a2})) has been postulated as the driving force for the dark energy, see footnotes \ref{DE},\ref{c_s} with some comments.
The model has been (successfully) confronted with observations, see \cite{Cai:2010uf,Sheykhi:2011xz, Sheykhi:2011nb,RozasFernandez:2011je,Cai:2012fq,Alcaniz:2012mh,Feng:2012gr,Malekjani:2012wc,Feng:2012wx} and many references therein, where it has been claimed that this proposal is consistent with all presently available data. Our comment here is that history of evolution of the universe may repeat itself by realizing the de Sitter behaviour twice in its history.  The $\qcd$-dynamics was responsible for the inflation in early universe, while   the QCD dynamics is responsible  for the dark energy in present epoch.

Our last comment is as follows. As we discussed at length in this paper, the  heart of the proposal is a fundamentally new type of energy which is not related to  any propagating degrees of freedom. 
Rather, this novel (non-dispersive) contribution to the  energy is formulated in terms of the tunnelling processes between topologically different but physically identical states. 
Our comment relevant for the present study is that this fundamentally new type of energy can be, in principle, studied in a laboratory
by measuring  the so-called topological Casimir Effect as suggested in  \cite{Cao:2013na,Zhitnitsky:2013hba}. The point is that if  the Maxwell theory is defined on a compact manifold there will be a  new contribution to the vacuum energy, in addition to the conventional Casimir energy.  This fundamentally new contribution  emerges as a result of tunnelling processes, rather than due to the conventional fluctuations of the propagating photons with two physical  polarizations. This effect does not occur for  the scalar field theory, in contrast with conventional Casimir effect which is operational for both: scalar as well as for Maxwell fields.  This extra energy computed in \cite{Cao:2013na,Zhitnitsky:2013hba} is the direct analog of the ``strange energy" which is the key player of the present paper. Furthermore, this fundamentally new type of energy can be also formulated in terms of auxiliary topological fields similar to $ [\phi(x), b(x)]$  fields introduced in this work, see \cite{Zhitnitsky:2013hba} for the details. In fact, the proposal  \cite{Cao:2013na,Zhitnitsky:2013hba} has been motivated in  an attempt to test the nature of the ``strange energy" as the critical element of the present studies\footnote{the idea to test some intriguing vacuum properties relevant for cosmology in a laboratory is not a very new idea. It has been advocated by Grisha Volovik  for years,   see recent review \cite{Volovik:2011dy} and references therein.}.

\section*{Acknowledgements}
I am thankful to Joe Polchniski for useful discussions  we had  while  he was visiting  Vancouver. 
This research was supported in part by the Natural Sciences and Engineering Research Council of Canada.
   
 \appendix
 \section{Linear correction $\sim H$ in deformed QCD. }\label{H}
 The main goal of this Appendix is to argue that the linear correction $\sim H$ indeed emerges in the deformed QCD model when the system is considered in the de Sitter background. Unfortunately,   the conventional computation scheme to carry out  a proper  computation as outlined in  steps 1-4   in Section \ref{odd}  is not   feasible due to the challenging technical problems.  Therefore, we use few additional simplified assumptions formulated below when the computations can be explicitly performed. First,  we assume that the  changes which occur in  the system  due to the curved background can be expressed in terms of  the same effective Lagrangian (\ref{b-action}) with the same auxiliary topological  fields $a (\mathbf{x}), b (\mathbf{x})$ as before, but written in a covariant way, without any additional terms. Second, due to some technical simplifications we can estimate  a correction to the energy due to the background field  at $\theta\neq 0$ which is  proportional to 
 $\theta^2 \left(1+{\cal{O}}(H)\right)$. We assume that the $\theta$ dependence (\ref{E_vac}) is not modified by the background. Therefore,  entire modification due to the background can be reconstructed for any $\theta$, including $\theta=0$.
In this case  the correction to the energy assumes the form $E_{\mathrm{vac}}(\theta) =-\frac{N\zeta}{L}\left[1+{\cal{O}}(H)\right]\cos\left(\frac{\theta}{N} \right) $ with a calculable coefficient in front of  $H$.

First, we explain  our approach  with estimations of the $\theta$ dependent portion of the vacuum energy
in flat space in section \ref{flat}. Our simplified procedure (which can be easily generalized to a curved background) leads to a parametrically correct expression given by eq. (\ref{E_vac}). It encourages us to   use  the same approximate method to estimate the $\theta$ dependent portion of the vacuum energy in    a curved background where we indeed observe the emergence of the linear correction $\sim H$,  see section \ref{deSitter}. The corresponding  linear correction $\sim H$ is interpreted  in section \ref{meaning} as a result of mixture of the gravitational background with topological  auxiliary field. With this interpretation we further elaborate on analogy with Aharonov -Casher effect mentioned in Introduction. This analogy now can be formulated in much more precise and specific way.  
Finally, in section \ref{QCD} we make few comments on  application of these results to   strongly coupled $\qcd$.
     \subsection{Simplified treatment of the ``strange energy" in  flat geometry at $\theta\neq 0$}\label{flat}
     The $\theta$ dependent portion of the vacuum energy in our system is known exactly, and it is given by (\ref{E_vac},\ref{zeta}).
     Furthermore, this ``strange" energy which can not be associated with any propagating degrees of freedom can be expressed in terms of  a correlation function (\ref{YM_top}) expressed in terms of auxiliary topological fields, see  \cite{Zhitnitsky:2013hs} 
    with all   technical details.  Our goal here is to reproduce this formula using a very simplified procedure which can be generalized to a curved background, when a corresponding exact formula is not known as we discussed in section \ref{odd}. 
     
   The action for the scalar auxiliary topological $b(\mathbf{x}), a(\mathbf{x}) $ fields   for our purposes  can be approximated  as follows 
      \be
\label{ba}
S [b, a] =
 \frac{-i }{4 \pi N}  \int_{\mathbb{R}^{3}}  d^{3}x     b(\mathbf{x})\vec{\nabla}^2 a (\mathbf{x}) 
	  - \zeta N \int_{\mathbb{R}^{3}}  d^{3}x  \cos \left( \frac{\theta-b(\mathbf{x})}{N} \right), 
		 \ee
     where we neglected the  fluctuations of massive physical scalar $\bm{\sigma}$ field by putting  $\bm{\sigma} =0$ in eq. (\ref{b-action}). Conventional way to compute the ``strange" energy in terms of the auxiliary fields is to integrate out  the $b(\mathbf{x})$   field, compute the corresponding correlation function at zero momentum transfer,  and express the vacuum energy in terms of this correlation function as discussed in details in ref. \cite{Zhitnitsky:2013hs}. 
     
  The corresponding  computational procedure  in a curved background is a very challenging problem. Therefore, we use the following simplified procedure for our estimates. We integrate out $b(\mathbf{x})$ field at $\theta\neq 0$ assuming that the fluctuations are small
    and  keeping the quadratic term in $\cos$ expansion, i.e. we consider the quadratic action 
        \be
\label{ba0}
S [b, a] =
 \frac{-i }{4 \pi N}  \int_{\mathbb{R}^{3}}  d^{3}x     b(\mathbf{x})\vec{\nabla}^2 a (\mathbf{x}) 
	  +\frac{ \zeta}{ 2N} \int_{\mathbb{R}^{3}}  d^{3}x   \left[{\theta-b(\mathbf{x})} \right]^2, 
		 \ee
which is known to reproduce all essential features of the system, such as topological susceptibility, see     ref. \cite{Zhitnitsky:2013hs} with details.
    As the  $b(\mathbf{x})$  field has  no  kinetic term, it is expressed in terms of $a (\mathbf{x}) $ field as follows
    \be
    \frac{\delta L  [b, a]}{\delta b(\mathbf{x})}=0 ~~~\rightarrow~~~ b(\mathbf{x})=\theta +\frac{i}{4\pi\zeta} \vec{\nabla}^2 a (\mathbf{x}). 
    \ee 
    We substitute this expression for $b(\mathbf{x})$ to eq. (\ref{ba0}) to arrive to the following effective action which determines the dynamics of the topological fields
       \be
\label{ba1}
S [b, a] =
 \frac{-i \theta}{4 \pi N}  \int_{\mathbb{R}^{3}}  d^{3}x \left[     \vec{\nabla}^2 a (\mathbf{x}) \right]
	  +\frac{1}{2\zeta N} \frac{1}{(4\pi)^2}\int_{\mathbb{R}^{3}}  d^{3}x \left[ a (\mathbf{x})     \vec{\nabla}^2  \vec{\nabla}^2 a (\mathbf{x})  \right].  
		 \ee 
  In our exact treatment in   ref. \cite{Zhitnitsky:2013hs} at $\theta=0$ we computed the corresponding Green's function, the topological susceptibility and the ``strange energy"   which follows from (\ref{ba1}) at $\theta=0$ when the first term in (\ref{ba1}) identically vanishes. We reproduced all  previous  results obtained  without even mentioning the auxiliary topological $b(\mathbf{x}), a(\mathbf{x}) $ fields.  As the corresponding direct computational scheme  outlined in section \ref{odd} represents a very challenging technical problem for a curved space background we shall use a simplified procedure for the  
    estimation which can be generalized to a curved background.  What is also important  is that all relevant elements of the system in this estimate can be understood in a simple and intuitive way such that the nature of the ``strange energy" becomes less mysterious.

  The idea is to compute the portion of the ``strange energy"  entering in combination  with $\theta$ parameter in the expansion $E_{\rm vac}(\theta)$.  
  Therefore, we shall only concentrate on the first term proportional to $\theta$  in eq. (\ref{ba1}) in this section to collect the terms proportional to $\theta^2$.
  To proceed with our  task we first 
  remind an exact formula for the vacuum expectation value for the topological density operator which directly follows from the definition   (\ref{thetaincluded}), 
  \be
  \label{q}
  \la iq \ra =\frac{1}{VL}\frac{\partial S(\theta)}{\partial \theta},
    \ee 
    where $VL$ is the 4-volume.   In deformed QCD model the corresponding expression for the vacuum energy $E_{\rm vac}(\theta)$  is known (\ref{E_vac}). Therefore   the expectation value for the topological density 
  can be represented as follows
    \be
    \label{q2}
    \la q(\mathbf{x})\ra =-i\frac{\zeta}{L}\cdot \sin \left( \frac{\theta}{N}\right), ~~~~ S(\theta)\equiv VL\cdot E_{\rm vac}(\theta), ~~~ E_{\rm vac}(\theta)=-\frac{N\zeta}{L}\cdot \cos \left( \frac{\theta}{N}\right). 
    \ee
    We note that the expectation value $ \la q(\mathbf{x})\ra$ (not the operator $q(\mathbf{x})$  itself) is complex  as it should as we are working in the  Euclidean space-time when a complex phase appears in the path integral formulation.  The same imaginary expectation value is known to occur in exactly solvable 2d QED, see e.g \cite{Zhitnitsky:2013hba}  with references on the original results.   In Minkowski space-time 
    $ \la q(\mathbf{x})\ra$ assumes a real value proportional to $\theta$ at small $\theta$. The expectation value $\la q(\mathbf{x})\ra$ has dimension four as the topological charge $Q=\int d^4 x q(x)$ representing   a specific  configuration of monopoles and anti-monopoles is a dimensionless  number.        The expectation value $\la q(\mathbf{x})\ra$  vanishes at $\theta=0$ as it should because the equal number of monopoles and anti-monopoles contribute to $\la q(\mathbf{x})\ra$ with equal weight, while for $\theta\neq 0$ the monopole's distribution is asymmetric leading to a non-vanishing expectation value  (\ref{q2}). 
   
    Now we want to interpret the known results (\ref{q2}) at small $\theta\ll 1$ in a simple intuitive way. This interpretation will play a key role in our discussions on  generalization of the    system to a curved background considered below  when exact formulae are not known.
  \exclude{ In deformed QCD model the expression for the topological density operator in terms of auxiliary topological field is determined by  eq. (\ref{YM_top}) and it is given by 
     \be
\label{q3}
q(\mathbf{x})=-\frac{1}{4 \pi NL}\left(\vec{\nabla}^2 a (\mathbf{x})\right)   .
\ee
Our normalization of the auxiliary topological field        $a (\mathbf{x})$ is such that every single monopole (anti-monopole) with asymptotic behaviour $a (\mathbf{x})=\pm1/ |\mathbf{x}|$ contributes $\pm1/N$ amount to the total topological charge of the  configuration.  
 The corresponding asymmetry at small $\theta$ can be easily extracted from  (\ref{q2}) by representing $ \la q(\mathbf{x})\ra$ as follows
\be
\label{q4}
 \la q(\mathbf{x})\ra = -\frac{\zeta}{L}\frac{\left[e^{i \frac{\theta}{N}}-e^{-i \frac{\theta}{N}}\right]}{2}\simeq -\frac{i\zeta}{2L}
 \left({\frac{\theta}{N}}+{ \frac{\theta}{N}}\right),~~~~~~~~~ \theta\ll 1,
\ee
where two terms in the brackets in eq. (\ref{q4}) are due to increasing monopole's density on a small amount $\sim \theta/N$ and decreasing the anti-monopole's density on the  same amount. 
}
First, we  consider a single monopole's contribution to the action  (\ref{ba1}) with $a (\mathbf{x})=1/ |\mathbf{x}|$.      As we  intend  in our simplified treatment to  estimate   only an additional  contribution proportional  to $\theta$ we  
limit ourself by studying the first term in action (\ref{ba1}) proportional to $\theta$.   A single monopole contributes  to the $\theta$ dependent portion of the action as follows,
\be
\label{q5}
\Delta S_{\rm (single~ mon.)}=
 \frac{-i \theta}{4 \pi N}  \int_{\mathbb{R}^{3}}  d^{3}x \left[     \vec{\nabla}^2 a (\mathbf{x}) \right]=\frac{i\theta}{N}.
\ee
Now,  we should multiply this  amount to the topological  density $(-\frac{i\zeta}{L})\cdot\left(\frac{\theta}{N}\right)$ from eq. (\ref{q2}) at small $\theta$. It represents the difference between the densities of monopoles and anti-monopoles which contribute to (\ref{q5}) with the opposite signs.
Finally one should multiply the obtained result to the total volume $(LV)$ and $N$ to account for all types of monopoles  in the entire 4-volume.
As a result of these multiplications we arrive to the following  order of magnitude estimate for the extra action due to non-vanishing $\theta$  
\be
\label{q6}
\Delta S_{\rm total}\simeq \left(\frac{i\theta}{N}\right)\cdot \left(\frac{-i\zeta}{L}\right)\cdot \left(\frac{\theta}{N}\right)\cdot (LV)\cdot N\simeq 
 \frac{\theta^2\zeta V}{N}.
\ee
  This represents a  parametrically correct  estimate  consistent with exact result    (\ref{q2}) for  small $\theta\ll 1$.
  The key observation here is that the system is gapped, but  the auxiliary topological field $a (\mathbf{x})$ is not screened.  In other words,   the auxiliary topological field $a (\mathbf{x})$  is effectively long ranged as discussed in great details in section \ref{BF-section} and specifically after eq. (\ref{K_top}). This is precisely the source  for non-vanishing  contribution to the action (\ref{q5}) from a single pseudo-particle with asymptotic behaviour $a (\mathbf{x})=1/ |\mathbf{x}|$ in a  plasma with a finite    Debye screening length.    Such a behaviour of the system should be contrasted with the well-known 3d Polyakov's model where a similar monopole's potential is screened, the contact term vanishes, and all effects (related to the  $\theta$ parameter) disappear, see few additional comments   in section \ref{meaning}. 
     
 \subsection{Corrections to the    ``strange energy" in de Sitter background}\label{deSitter}
 The main goal of this subsection is to generalize the simplified  estimates (\ref{q5}) and (\ref{q6}), which represent the $\theta$ dependent portion of the ``strange energy", on a time dependent  background  parametrized by the  Hubble constant $H$. We do not want to destroy the weak coupling  regime of the deformed QCD. Therefore, we do not change   parameter $L$ which is the length of the compactified 4-th dimension  in this system. Instead, we want to  model the de Sitter behaviour by modifying  the geometry $\mathbb{R}^{3}$ of  the system defined  by the topological action (\ref{ba}).  With this purpose we consider three dimensional Minkowski space $\mathbb{R}^{(1,2)}$ with Lorentzian signature instead of the original Euclidean signature $\mathbb{R}^{3}$ which enters (\ref{ba}). After that, one can introduce a  scale factor ${\rm a}(t)$ which models the expansion of the universe. The next  conventional step is to use the conformal time $\eta$ instead of physical time $t$ 
 \be
 d\eta=\frac{dt}{{\rm a}(t)}.
 \ee
 To simplify analysis we concentrate on the de Sitter behaviour with the following properties
 \be
 \frac{\dot{\rm a}(t)}{{\rm a}(t)}\equiv H, ~~ {\rm a}(\eta)=-\frac{1}{H\eta}, ~~ H={\rm const}.
 \ee
   Furthermore, we assume  that $H$ is  much smaller than all other scales of the problem. 
As  the next step  we follow  a conventional procedure when the scale factor ${\rm a}(t)$  can be removed from the action by introducing    $\bar{a}$ and $\bar{b}$ fields and rescaling the dimensional parameter $\zeta$ of the system as follows:
 \be
 \label{variables}
   \bar{a} \equiv  {\rm a}(t) a, ~~  \bar{b}\equiv  b, ~~~\bar{\zeta}\equiv  {\rm a}^3(t) \zeta .
 \ee  
Our study of the ``strange energy" is formulated using the  Euclidean signature in terms of pseudoparticles  (monopoles)
which describe the tunnelling events, see text after eq.(\ref{zeta}). Therefore, once parameter $H$ is introduced into the system we return to the metric with the Euclidean signature using conventional analytical continuation. 
As a result of this    procedure we  arrive to  the following action in terms of new $\bar{b}(\mathbf{x},\eta)$ and $\bar{a}(\mathbf{x},\eta)$  fields, 
     \be
\label{ba2}
S [\bar{b}, \bar{a}] =
 \frac{-i }{4 \pi N}  \int  d^{2}\mathbf{x} d\eta   \bar{b}(\mathbf{x},\eta)  \left[\vec{\nabla}^2 \bar{a} (\mathbf{x},\eta)+\frac{2}{\eta}\frac{\partial\bar{a} (\mathbf{x},\eta)}{\partial \eta}  \right]
	  +\frac{ \bar{\zeta}}{2N}  \int  d^{2}\mathbf{x} d\eta    \left[{\theta-\bar{b}(\mathbf{x},\eta)} \right]^2, ~~~ \vec{\nabla}^2
 \equiv \frac{\partial^2}{\partial \mathbf{x}^2}+\frac{\partial^2}{\partial \eta^2}.~~~~
		 \ee
 In formula (\ref{ba2})  
 we use   $\mathbf{x}$    for two expanding coordinates, while  $\eta$ in eq. (\ref{ba2}) represents the Euclidean conformal time. 
  
  We should remark 
   here that the both technical elements employed in deriving (\ref{ba2})  describing the action  with  Euclidean signature in curved space   are conventional technical tools, see e.g. \cite{mukhanov}. In particular, in case of a   massive field with a canonical kinetic term the problem is reduced to a conventional QFT in a flat background with the only new element is that a time-dependent effective mass appears in the description, 
   \be
   \label{mass}
   m^2_{\rm eff}\equiv {\rm a}^2m^2-\frac{1}{\rm a}\frac{\partial^2 \rm a}{\partial \eta^2}= {\rm a}^2m^2 -\frac{2}{\eta^2}.
      \ee
 Our   original topological action (\ref{ba0}) does not have a   canonical kinetic terms as the system does not describe any propagating degrees of freedom. As a result of this difference with canonical case (\ref{mass}) the only new element which emerges  in   eq. (\ref{ba2}) is an extra term
  $\left[\frac{2 \bar{b}}{\eta}\frac{\partial\bar{a} (\mathbf{x},\eta)}{\partial \eta} \right]$. 
  
 As our system (\ref{ba2}) is formulated in the same terms as in  original formulation (\ref{ba0}) we simply repeat all steps leading to the simplified  estimates (\ref{q5}) and (\ref{q6})  taking into account an additional term in squared brackets (\ref{ba2}). As a result of this procedure, we arrive to the following extra contribution from a single pseudo-particle   with the asymptotic behaviour 
 $ \bar{a} (\mathbf{x},\eta)=1/\sqrt{\mathbf{x}^2+\eta^2}$
  \be
\label{q7}
\Delta S_{\rm (single~ mon.)}=
 \frac{-i \theta}{4 \pi N}  \int  d^{2}\mathbf{x} d\eta \left[\vec{\nabla}^2 \bar{a} (\mathbf{x},\eta)+\frac{2}{\eta}\frac{\partial\bar{a} (\mathbf{x},\eta)}{\partial \eta}  \right]\nonumber\\
 =\frac{i\theta}{N}\left[1+\int\frac{d\eta}{\eta}\right]= \frac{i\theta}{N}\left[1- H\int  dt\right]= \frac{i\theta}{N}\left[1- {\cal{O}}\left(\frac{H}{\sqrt[3]{\zeta}}\right)\right],
\ee
 where in last term we returned to the physical time $t$ variable instead of the conformal time $\eta$. Furthermore, we cutoff the integral $\int dt$ at the scale $1/\zeta$ which is the only physical scale of the problem, and roughly corresponds to a typical time scale of  the tunnelling events.  The corresponding correction $\sim H$ will also enter formula (\ref{q6}) in front of $\theta^2$ term.
 Assuming that the $\theta$ dependence is not altered by a curved background we formulate our final estimate in the same form
 as presented  in  section \ref{odd}
  \be
 \label{power1}
 E_{\rm vac}(H)= -\frac{N\zeta}{L}\left[1- {\cal{O}}\left(\frac{H}{\sqrt[3]{\zeta}}\right)\right].
 \ee
 Formula (\ref{power1}) represents an extra contribution to 
  ``strange energy"  due to the  tunnelling events in expanding background parametrized by the Hubble constant $H$. In the Euclidean formulation the same  extra energy describes the  variation of  the monopole density as a result of expansion.

 \subsection{Interpretation}\label{meaning}
 Our goal here is to explain a highly nontrivial phenomenon represented by eq. (\ref{power1}) with a linear correction $\sim H$ which  naively  contradicts to   a  conventional viewpoint  that the correction must be quadratic, see section \ref{odd}.
  One can explicitly see from eq. (\ref{q7}) that the crucial element for the linear correction to emerge is the presence of the long ranged  topological field which     mixes  with the background  
  represented by the Hubble parameter $H$. Indeed, if in eq. (\ref{q7})  instead of $ \bar{a} \sim r^{-1}$ we would use a  screened massive  field   i.e. $\varphi (r)\sim \exp (-m r)/r $ we would obviously get the vanishing contribution from the large distances. 
  This is in fact exactly the case for 3d Polyakov's model which is known to produce a vanishing contact term  as all physical results are $\theta$ independent in that model in large volume limit. Therefore, the origin of linear correction $\sim H$ lies in  understanding of the 
long range behaviour of the topological field in a gapped system.    Formally, this long range behaviour is formulated in terms of the contact term proportional to the $ \delta(\bold{x}) $ function (\ref{YM_top}), or what is the same in terms of the massless pole (\ref{K_top}).

Normally, a pole at zero mass corresponds to a massless gauge boson. Or it might be a result of spontaneous symmetry breaking effect. However, we do not have any physical massless states in the system. What is a symmetry which could be responsible for behaviour (\ref{K_top})?  Furthermore,  this pole must have a residue with a Òwrong signÓ such that it can not be identified with any physical propagating massless degree of freedom.  In weakly coupled ``deformed QCD" the contact term is saturated by monopoles which describe the tunnelling between physically identical, but  topological  different   winding  $| n\ra$  states.  Therefore, one can interpret the symmetry which is  responsible for such a behaviour as the 
invariance under the   large gauge transformations   as argued in \cite{Zhitnitsky:2011aa}.
It is important  to emphasize that while the operator ${\cal{T}}$ is formally constructed as an operator of gauge transformations, this operator does change the state   as a result of global effect, i.e. ${\cal T} | n\ra= | n+1\ra$. Therefore, one should treat ${\cal{T}}$ as ``improper" gauge transformation (the ``large gauge transformation"). Still, ${\cal{T}}$ commutes with the hamiltonian $[{\cal{T}}, H]=0$. 
Precisely this feature (\ref{K_top}) with a topologically protected massless pole is eventually responsible for the linear correction (\ref{power1}) to the ``strange energy" as any massive physical states can not produce such type of effects.

  It is quite instructive to present some analogy with a system  which is realized in nature and which exhibits   similar properties. 
    While  there are many (topologically ordered) systems which demonstrate similar features, we concentrate on the well known  Aharonov -Casher effect as formulated in \cite{Reznik:1989tm}.
The relevant part of this work can be stated as follows. If one inserts an external charge into superconductor when the electric field is exponentially suppressed $\sim \exp(-r/\lambda)$ with $\lambda $ being the penetration depth,  a neutral magnetic fluxon will be still sensitive to an inserted external charge at arbitrary large distance. The effect is pure topological and non-local in nature.  The crucial element why this phenomenon occurs in spite of the fact that the system is gapped is very similar to our case. First of all, it is the presence of different topological states  $u_n$ (number of Cooper pairs) in the system and ``tunnelling" between them  (non-vanishing matrix elements  between $u_n$ and $u_{n+1}$ states) as described in  \cite{Reznik:1989tm}. Those states are  analogous  to the topological sectors $|n\ra$ in our work.
As a result of the ``tunnelling", an appropriate ground state $U(\theta)$ must be constructed as discussed  in \cite{Reznik:1989tm}, analogous to the  $|\theta\ra$ vacuum construction in gauge theories.
This state $U(\theta)$ is an eigenstate of the so-called ``modular operator" which commutes with the hamiltonian. In our work an analogous role   plays the large gauge transformation operator $\cal{T}$ such that $ {\cal{T}}|\theta\ra=\exp(-i \theta )| \theta\ra$. An explicit construction of the operator  $\cal{T}$ is known: it is non-local operator   similar to non-local ``modular operator" from ref. \cite{Reznik:1989tm}, see Appendix  in ref. \cite{Zhitnitsky:2011aa} for some technical details in the given context. 
\exclude{The crucial element of the construction of ref.\cite{Reznik:1989tm} is that the induced charges in presence of the gap can not screen the ``modular charge" as a result of commuting the ``modular operator" with hamiltonian.  This eventually leads to a non-vanishing effect, i.e.  collecting a non-zero phase at arbitrary large distance. 
 We do not have a luxury to resolve the problem using the hamiltonian description in  strongly coupled four dimensional QCD. However,  one can argue that  the role of ``modular operator"
is played by large gauge transformation operator $\cal{T}$ which also commutes with the hamiltonian $[{\cal T},H]=0$, such that} Our system is  transparent to   topologically nontrivial pure gauge configurations, similar to transparency of the superconductor to the ``modular electric field" from ref. \cite{Reznik:1989tm}. Such a behaviour of our  system can be thought as a non-local topological effect  similar to the   non-local Aharonov -Casher effect as formulated in \cite{Reznik:1989tm}.   

We should emphasize again that the are no any 
physical propagating massless degrees of freedom   in the system. 
The description of the system in terms of auxiliary topological fields saturating the correlation function   (\ref{K_top}) with seemingly a massless pole is not a mandatory, but a matter of convenience. Similarly, the description of a topologically ordered phase in condensed matter physics in terms of Chern Simons effective Lagrangian is a matter of convenience rather than a necessity.  In fact all relevant features of topologically ordered phases (such as braiding phases or degeneracy of the ground state) have been originally  established without   any auxiliary Chern Simons fields. The same comment also applies to  our case when the {\it inflaton} is an auxiliary, not dynamical,  topological field which effectively describes the dynamics of the topological sectors of the gauge system in expanding universe. 
In principle, one could follow streps 1-4 from section \ref{odd} to compute the correction (\ref{power1}) without any auxiliary fields.  However, our estimate (\ref{power1}) demonstrates the convenience of  the topological field which we identify with {\it inflaton}.  
Furthermore, the formal similarities with Aharonov -Casher effect presented above   makes the analogy mentioned  in the Introduction on the nature of the {\it inflaton} field much more specific and precise.

 \subsection{Generalization to four dimensional $\qcd$}\label{QCD}
 In previous subsections we have argued, using  the weakly coupled deformed QCD model, that the correction to the  ``strange energy"  could  demonstrate a  linear $\sim H$ scaling rather than naively expected $\sim H^2$ scaling. In strongly coupled $\qcd$ we can not use the same technique as our semiclassical computation is not justified.
 However, as claimed in  \cite{Yaffe:2008} the transition from weakly coupled deformed QCD to strongly coupled QCD must be smooth, without any phase transitions on the way. Therefore, one should expect that the same linear scaling should hold in strongly coupled regime as well. This is precisely the key assumption of  sections \ref{odd}, \ref{inflation}, and in fact, entire framework advocated by  the present paper. Below we present an additional argument further supporting this assumption.
 
 The argument is based on the observation that the crucial element leading to the linear $\sim H$ scaling is the 
 presence of an effectively massless  auxiliary topological field expressed by the correlation function  (\ref{K_top}). 
 The assumption  of the continuity in the passage from the weakly coupled to the strongly coupled regime is formulated 
 in terms of the topological fields as a prescription that the only dimensional parameter $\zeta/L$ from weakly coupled deformed QCD becomes $\Lbar^4$   in strongly coupled $\qcd$. The inflaton  field $a(\mathbf{x})$ from the  deformed QCD construction  is replaced by $\phi (x) $ field in strongly coupled $\qcd$. The new inflaton field $\phi (x) $  should be  identified with the  longitudinal component of $K_{\mu}(x)\sim\partial_{\mu}\phi (x)$ such that topological density operator is $q(x)\sim \Box\phi(x)$ assumes the same form as  $ q(\mathbf{x}) \sim  \vec{\nabla}^2 a (\mathbf{x})$ from the deformed QCD model.    Another auxiliary field $b(\mathbf{x})$ always enters the effective description along with $\theta$ parameter in the  combination $\theta\rightarrow [\theta -b(\mathbf{x})]$.   In strongly coupled $\qcd$ the $b(x)$   must keep its transformation properties.  As a result of these replacements we arrive to the following effective 
 low action for the topological $[b, \phi] $ fields,   
      \be
\label{phi1}
S [b, \phi] =
 -i \int_{\mathbb{R}^{4}}  d^{4}x     b( {x}){\Box} \phi ({x}) 
	  +\frac{1}{2} { \Lbar^4} \int_{\mathbb{R}^{4}}  d^{4}x   \left[ \theta-b ({x}) \right]^2. 
		 \ee
 This action  replaces eq. (\ref{ba0}) for weakly coupled deformed QCD. There is a fundamental difference between eq. (\ref{ba0}) and (\ref{phi1}). In former  case the corresponding action has been derived in \cite{Zhitnitsky:2013hs} from the first principles in the weakly coupled gauge theory, while in the later case it has been reconstructed above assuming the  continuity in the passage from the weakly coupled to the strongly coupled regime. 
 
 Nevertheless, one can argue that (\ref{phi1}) represents a correct description of the low energy physics. In particular, it saturates the contact term in the topological susceptibility (\ref{K}), (\ref{K1}).
 Indeed, one can integrate $b(x)$ field in eq. (\ref{phi1}) such that the effective action for $\theta=0$ becomes
       \be
\label{phi2}
S [\phi] =  \Lbar^{-4}\cdot
	  \frac{ 1}{2} \int_{\mathbb{R}^{4}}  d^{4}x     \phi( {x}){\Box}\Box \phi ({x}). 
		 \ee
		Such an  effective action written in the form  $\int d^4x q^2(x)\sim \int d^4x (\partial_{\mu}K^{\mu})^2$ has been, in fact,  postulated by Veneziano \cite{ven,vendiv} as the key element in the resolution of the $U(1)_A$ problem.  		
		The relevant correlation functions can be explicitly evaluated now from eq. (\ref{phi2})  with  the results
   \be
  \label{phi3}
  && \< q({x}) q({0}) \> \sim  \la \Box \phi ({x}) , \Box \phi (0)\ra \sim 
   \frac{\int {\cal{D}[\phi]} e^{-S(\phi)} \Box \phi ({x}) \Box \phi ({0})}{\int {\cal{D}[\phi]} e^{-S(\phi)}}    \sim      \Lbar^4 \cdot \delta^4(x)\, \\
   && \lim_{k\rightarrow 0} \int d^4x e^{ikx} \la K_{\mu}(x) , K_{\nu}(0)\ra \sim  
      \lim_{k\rightarrow 0} \int d^4x e^{ikx} \la \partial_{\mu}\phi(x) , \partial_{\nu}\phi(0)\ra\sim  
     \Lbar^4 \cdot\frac{k_{\mu}k_{\nu}}{k^4}. \nonumber
   \ee 
  The main  point here is that the   effective action (\ref{phi1}) does reproduce all relevant elements  (\ref{K}), (\ref{K1}) which are known to be present in strongly coupled QCD. 
   
   Now we can repeat all  steps we employed in previous subsections \ref{flat},  \ref{deSitter} and \ref{meaning} to 
  generalize our system to a curved background characterized by the Hubble parameter $H$.   In particular, we introduce the rescaled field, similar to eq. (\ref{variables}) as follows, 
  \be
 \label{variables1}
   \bar{\phi} \equiv  {\rm a^2}(t) \phi, ~~  \bar{b}\equiv  b, ~~~\bar{\Lambda}_{\qcd}\equiv  {\rm a}(t) \Lbar .
 \ee  
  In terms of the rescaled variables the action in the  Euclidean signature assumes the following form
       \be
\label{phi4}
S [b, \phi] =
 -i \int  d^{3}\mathbf{x} d\eta  \cdot  \bar{b}( \mathbf{x}, \eta)\left[{\Box} \bar{\phi} ({\mathbf{x},\eta})+\frac{4}{\eta}\frac{\partial \bar{\phi}(\mathbf{x},\eta)}{\partial \eta} \right]
	  +\frac{1}{2}\bar{\Lambda}^4_{\qcd} \int d^{3}\mathbf{x} d\eta  \cdot    \left[ \theta-\bar{b} (\mathbf{x},\eta) \right]^2, ~~  \Box\equiv \frac{\partial^2}{\partial \mathbf{x}^2}+\frac{\partial^2}{\partial \eta^2}, ~~~~~
		 \ee
 where we use   $\mathbf{x}$    for three  expanding coordinates while $\eta$ is the Euclidean conformal time. The structure of this action is very much the same as the action for weakly coupled gauge theory given by eq.(\ref{ba2}). 
 The extra term $\sim \left(\frac{b}{\eta}\right)\cdot\left(\frac{\partial \bar{\phi} }{\partial \eta}\right)$ describing the mixture of the inflaton field with a curved backgroun also assumes the  same strcture.  However, we can not proceed with estimations,  similar to eq. (\ref{q7}), because there are no well-defined weakly interacting pseudo-particles (similar to monopole-instanton) in strongly coupled QCD. Nevertheless, it is naturally to expect that the correction to the ``strange energy" due to the expanding universe, will also exhibit the linear scaling, similar to eq.(\ref{power1}),   i.e.
 \be
 \label{power2}
 E_{\rm vac}(H)= -N^2\Lbar^4\left[1- {\cal{O}}\left(\frac{H}{\Lbar }\right)\right].
 \ee 
 This expectation is based on observation that the key element leading to the linear   correction in eq.(\ref{power1}) is the presence of 
 the long ranged   topological field as explained in section \ref{meaning}. This feature is obviously present in strongly coupled regime (\ref{phi3}), in close analogy with  the corresponding  expressions (\ref{YM_top}) and (\ref{K_top}) derived for the weakly coupled deformed QCD. In both cases the linear term can be interpreted as the result of mixture  of the topological {\it inflaton}  field  $\left(\frac{  \partial \bar{\phi} }{\partial \eta}\right)$ with the curved background represented by $\left(\frac{1}{\eta}\right)$.

 \section{Induced $\theta_{ind}(x)$,  the helical instability, and the linear $\sim H$ scaling in heavy ion collisions.}\label{ions}
 The goal of this  Appendix is to present some analogy with a system which has precisely the structure (\ref{coup}). The structure (\ref{coup}) has emerged in the context of the present work as  the coupling  between auxiliary  field $b(x)$  and physical gauge fields,  and it was   was the crucial element in our presentation on possible reheating mechanism  within $\qcd$- {\it inflation } scenario  in section \ref{reheating}. 
 The same structure also emerges in context of heavy ion physics. To be more specific, it has been suggested a while ago
 \cite{Kharzeev:1998kz,Kharzeev:1999cz,Halperin:1998gx,Fugleberg:1998kk,Buckley:1999mv} that the so-called induced theta vacua $|\theta_{\rm ind}\ra$ can be created in heavy ion collisions. This direction of studies became  very active area of research after an appropriate observational signature has been suggested \cite{Kharzeev:2004ey}  and theoretical computations of the effect have been put on a solid theoretical ground  
 \cite{Kharzeev:2007tn}, see also related papers \cite{Kharzeev:2007jp,Fukushima:2008xe,Zhitnitsky:2012im}
 with specific applications to heavy ion collisions, and also
 review paper \cite{Kharzeev:2009fn} which covers some  recent theoretical development.   
 The experimental studies at the relativistic heavy ion collider (RHIC), Brookhaven \cite{Abelev:2009tx} and more recently, at the LHC \cite{Abelev:2012pa} apparently support the basic picture advocated in \cite{Kharzeev:2004ey,Kharzeev:2007tn,Kharzeev:2007jp,Fukushima:2008xe,Zhitnitsky:2012im,Kharzeev:2009fn}. 
 
 The basic idea advocated in \cite{Kharzeev:2004ey,Kharzeev:2007tn,Kharzeev:2007jp,Fukushima:2008xe,Zhitnitsky:2012im,Kharzeev:2009fn} can be explained in few lines as follows. 
 Let us assume that an  effective   $\theta (\vec{x}, t)_{ind}\neq 0$  is induced as a result of some nonequlibrium dynamics as suggested in refs.
 \cite{Kharzeev:1998kz,Kharzeev:1999cz,Halperin:1998gx,Fugleberg:1998kk,Buckley:1999mv}. The  $\theta (\vec{x}, t)_{ind}$ parameter enters the effective lagrangian as follows, 
 \be
 \label{theta}
  {\cal L_{\theta}}(x)=- \frac{g^2}{64\pi^2} \theta_{ind}(x) \epsilon_{\mu\nu\rho\sigma} F^{a\mu\nu} F^{a\rho\sigma}(x)
  \ee
  which is very similar in spirit    to eq. (\ref{coup}) describing  the interaction   between auxiliary  field $b(x)$  and physical gauge fields. 
  In context of heavy ion collisions the  $\theta_{ind}(x)$  plays the same role as  $b(x)$ does  in present work. In both cases these fields are not dynamical, and in both cases they reflect the changes related to the variation of the environment (colliding nuclei versus expanding universe). In both cases these auxiliary fields code the information on modification  of the topological sectors 
  as a result of this variation. It is obvious that the typical scales are   very different in these two problems: in expanding universe the scale is determined by Hubble parameter $H$, while in heavy ion physics a typical scale is determined by  a correlation length of the  $\theta_{ind}(x)$ which is a  size of a nuclei $L$. What is important is that in both cases these scales are parametrically smaller than internal fast fluctuations, i.e. $H\ll \Lbar$ and $L^{-1}\ll \Lqcd$ scales correspondingly. It allows to treat $b(x)$ and $\theta_{ind}(x)$ as the external slow varying background fields when the effective Lagrangian approach (\ref{coup}) and (\ref{theta}) correspondingly, is justified.  
  
  In context of heavy ion collisions the interaction (\ref{theta}) has received a lot of attention because it implies that 
  the  local ${\cal{P}}$  and ${\cal{CP}}$
  invariance of QCD is broken on the scales where correlated $\theta (\vec{x}, t)_{ind}\neq 0$  is induced. As a result of this violation, one should expect a number of ${\cal{P}}$  and ${\cal{CP}}$ violating effects taking place in the region where $\theta (\vec{x}, t)_{ind}\neq 0$.  In particular, in the presence of an external magnetic field $\vec{B}$ or in case of the rotating  system  with angular velocity $\vec{\Omega}$ there will be induced electric   current directed along $\vec{B}$ or $\vec{\Omega}$ correspondingly, resulting in   separation
 of charges along those directions. This leads to a number of effects such as ``charge separation effect", ``chiral magnetic effect", 
 ``chiral vortical effect", etc, see review article \cite{Kharzeev:2009fn} for a short introduction into the field. Apparently, the corresponding effects have been observed at RHIC  \cite{Abelev:2009tx} and confirmed at the LHC energies \cite{Abelev:2012pa}.   QCD itself obviously does not break  ${\cal{P}}$  and ${\cal{CP}}$ invariance on the fundamental level. It implies that all these effects must be measured on event by event basis when $\theta_{ind}(x)$ parameter assumes a different value with a different sign in each given event. 
  
  In context of the present work it is important to notice that for a  time dependent and spatial-independent  $\theta (t)_{ind}\neq 0$ can be represented as non-vanishing axial chemical potential for a massless fermion $\psi$. Indeed, one can perform in the path integral a $U(1)_A$ chiral time-dependent transformation   to rotate away   the coupling (\ref{theta}). The corresponding interaction reapers in the form of a non-vanishing axial chemical potential $(\mu_L-\mu_R)\neq 0$. To be more precise, 
  \be
  \label{mu5}
  \psi\rightarrow \exp\left(i\frac{\theta (t)_{ind}}{2}\right) \psi, ~~~~~ \bar{\psi}\gamma_{\mu}D^{\mu}\psi\rightarrow 
 \bar{\psi}\gamma_{\mu}D^{\mu}\psi + (\mu_L-\mu_R) \bar{\psi}\gamma_0\gamma_5\psi, ~~~~~ \mu_5\equiv(\mu_L-\mu_R)\equiv {\dot\theta (t)_{ind}}, 
  \ee
  see also~\cite{Kharzeev:2009fn}
  for a physical interpretation of this  relation $(\mu_L-\mu_R)={\dot{\theta }(t)_{ind}}$.  One should comment here that the axial chemical potential $\mu_5$ does not correspond to any conserved charges, in contrast with conventional chemical potential $\mu$ which is related to the conservation of the baryon charge.  Nevertheless, $\mu_5$  can be used in computations assuming it is a slow varying function of time. 
  
  In context of our work when the typical fluctuations of $b(x)$  (playing  the role of $\theta (t)_{ind}$, as explained above) are of order of $H$, one can also identify $ |{\dot{\theta }(t)_{ind}}|\rightarrow |\dot{b}(t)|\sim H$ with local generation of $|\mu_5|\sim H$ on that scales. With these identification, one can use the recent studies \cite{Akamatsu:2013pjd} on computation of the helical instability in plasma with the result that the time scale of the plasma instability is  \cite{Akamatsu:2013pjd}:
  \be
  \label{tau}
  \tau_{\rm inst}\sim \frac{1}{\alpha_s^2\mu_5}.
  \ee
  With our identification $|\mu_5|\sim H$ in cosmological context we arrive to estimate (\ref{e-folds}). It has been also argued in  \cite{Akamatsu:2013pjd} that the fate of this instability is to reduce $\mu_5$ which itself is a source of this instability. In our cosmological context it implies that the fate of instability is to reduce the inflationary Hubble constant (\ref{a2}).  The corresponding inflationary energy (\ref{friedman-infl})  which is proportional to $H$ will be  transferred to the light particles, which is precisely the destiny of the reheating epoch. 
  
  We conclude the Appendix with one more  short remark  on analogy between heavy ion collisions characterized by $\theta (t)_{ind}$ and cosmology characterized by $b(x)$ in our framework. 
  The point is 
  that a linear dependence on $H$ as discussed in section \ref{odd}
  can be, in principle, tested in context of heavy ion collisions as discussed in \cite{Zhitnitsky:2012im}. The key element in studying 
  the local violation of the ${\cal{P}}$  and ${\cal{CP}}$
  invariance
   is that the typical correlation length for $\theta (t)_{ind}$ in heavy ion collisions context is played by a size of a nuclei $L$. When the size of a nuclei varies, the effect must scale as $L^{-1}$ which plays the role of the Hubble parameter $H$ in the cosmological context. Available experimental data apparently support $L^{-1}$ scaling law as the studies have been performed for a number of nuclei with different sizes: ${\rm Au^{79}, ~Pb^{82}, Cu^{29}}$,  see  \cite{Zhitnitsky:2012im} for the details.

\end{document}